\documentclass[prd,aps,preprint,tightenlines,nofootinbib,superscriptaddress]{revtex4}
\usepackage{mathrsfs}
\usepackage{amsfonts}
\usepackage{amsmath}
\usepackage{array}
\usepackage{verbatim}
\usepackage{epsfig}
\usepackage{graphicx}
\usepackage{hyperref}

\begin{document}
\title{Factorization and its Breaking in Dijet Single Transverse Spin Asymmetries in $pp$ Collisions}

\author{Xiaohui Liu}
\affiliation{Center of Advanced Quantum Studies, Department of Physics,
Beijing Normal University, Beijing 100875, China}

\author{Felix Ringer}
\affiliation{Nuclear Science Division, Lawrence Berkeley National
Laboratory, Berkeley, CA 94720, USA}

\author{Werner Vogelsang}
\affiliation{Institute for Theoretical Physics,
                Universit\"{a}t T\"{u}bingen,
                Auf der Morgenstelle 14,
                D-72076 T\"{u}bingen, Germany}
                
\author{Feng Yuan}
\affiliation{Nuclear Science Division, Lawrence Berkeley National
Laboratory, Berkeley, CA 94720, USA}

\begin{abstract}
We study factorization in single transverse spin asymmetries for dijet production in proton-proton collisions, by considering soft gluon radiation at one-loop order. We show that the associated transverse momentum dependent (TMD) factorization is valid at the leading logarithmic level. At next-to-leading-logarithmic (NLL) accuracy, however, we find that soft gluon radiation generates terms in the single transverse spin dependent cross section that differ from those known for the unpolarized case. As a consequence, these terms cannot be organized in terms of a spin independent soft factor in the factorization formula. We present leading logarithmic predictions for the single transverse spin dijet asymmetry for $pp$ collisions at RHIC, based on quark Sivers functions constrained by semi-inclusive deep inelastic scattering data. We hope that our results will contribute to a better understanding of TMD factorization breaking effects at NLL accuracy and beyond.
\end{abstract}
\maketitle

\section{Introduction}

There has been a strong interest in correlated dijet production in various hadronic collisions~\cite{Abazov:2004hm,Abelev:2007ii,Khachatryan:2011zj,daCosta:2011ni,Aad:2010bu,Chatrchyan:2011sx,Adamczyk:2013jei,Adamczyk:2017yhe,Aaboud:2019oop}, where the two jets are produced mainly in the back-to-back configuration in the transverse plane,
\begin{equation}
A+B\to {\rm Jet}_1+{\rm Jet}_2+X \ . \label{dijet}
\end{equation}
Here $A$ and $B$ represent the two incoming hadrons with momenta $P_A$ and $P_B$, respectively. The azimuthal angle between the two jets is defined as $\phi=\phi_1-\phi_2$ with $\phi_{1,2}$ being the azimuthal angles of the two jets. In the leading order naive parton picture, the Born diagram yields a delta function at $\phi=\pi$. One-loop gluon radiation will lead to a singular distribution around $\phi=\pi$. This divergence arises when the total transverse momentum of the dijet (imbalance) is much smaller than the individual jet momentum, $q_\perp=|\vec{k}_{1\perp}+\vec{k}_{2\perp}|\ll |k_{1\perp}|\sim |k_{2\perp}|\sim P_T$, where large logarithms appear at every order of the perturbative calculation. In the kinematic region $q_\perp\ll P_T$, the appropriate resummation method that needs to be applied is the so-called transverse momentum dependent (TMD) resummation or the Collins-Soper-Sterman (CSS) resummation~\cite{Collins:1984kg}. There have been several theoretical efforts to resum the large logarithms for this process~\cite{Banfi:2003jj,Banfi:2008qs,Hautmann:2008vd,Chiu:2012ir,Mueller:2013wwa,Sun:2014gfa,Sun:2015doa,Chien:2020hzh}. The differential cross section can be written as,
\begin{eqnarray}
\frac{d^4\sigma}
{d\Omega
d^2q_{\perp}}=\sum_{abcd}\sigma_0\left[\int\frac{d^2\vec{b}_\perp}{(2\pi)^2}
e^{i\vec{q}_\perp\cdot
\vec{b}_\perp}W_{ab\to cd}(x_1,x_2,b_\perp)+Y_{ab\to cd}\right] \ ,
\end{eqnarray}
where $d\Omega=dy_1 dy_2 d P_T^2$ represents the phase space of dijet production. Here $y_1$ and $y_2$ are the rapidities of the two jets, $P_T$ is the leading jet transverse momentum, and $q_\perp$ the imbalance transverse momentum between the two jets as defined above. Moreover, $\sigma_0$ is the overall normalization of the differential cross section. The first term on the right hand side, $W_{ab\to cd}$, contains the all order resummation and the second term, $Y_{ab\to cd}$, takes into account fixed order corrections. At next-to-leading logarithmic (NLL) order, the resummation for $W$ was conjectured to take the following form~\cite{Sun:2014gfa,Sun:2015doa}
\begin{eqnarray}
W_{ab\to cd}\left(x_1,x_2,b\right)&=&x_1\,f_a(x_1,\mu=b_0/b_\perp)
x_2\, f_b(x_2,\mu=b_0/b_\perp) e^{-S_{\rm Sud}(Q^2,b_\perp)} \nonumber\\
&\times& \textmd{Tr}\left[\mathbf{H}_{ab\to cd}
\mathrm{exp}\left[-\int_{b_0/b_\perp}^{Q}\frac{d
\mu}{\mu}\mathbf{\gamma}_{ab\to cd}^{s\dag}\right]\mathbf{S}_{ab\to cd}
\mathrm{exp}\left[-\int_{b_0/b_\perp}^{Q}\frac{d
\mu}{\mu}\mathbf{\gamma}_{ab\to cd}^{s}\right]\right]\ ,\label{resum}
\end{eqnarray}
for each partonic channel $ab\to cd$, where $Q^2=\hat s=x_1x_2S$, representing the hard momentum scale. In addition, we have $b_0=2e^{-\gamma_E}$, with $\gamma_E$ being the Euler constant. The $f_{a,b}(x,\mu)$ are the parton distributions for the incoming partons $a,b$, and $x_{1,2}=P_T\left(e^{\pm y_1}+e^{\pm y_2}\right)/\sqrt{S}$ are the fractions of the incoming hadrons' momenta carried by the partons. In the above equation, the hard and soft factors $\mathbf{H}$ and $\mathbf{S}$ are expressed as matrices in the color space of the partonic channel $ab\to cd$, and $\gamma_{ab\to cd}^s$ are the associated anomalous dimensions for the soft factor. The Sudakov form factor ${\cal S}_{\textrm{Sud}}$ resums the leading double logarithms and the universal sub-leading logarithms,
\begin{eqnarray}
S_{\rm Sud}(Q^2,b_\perp)=\int^{Q^2}_{b_0^2/b_\perp^2}\frac{d\mu^2}{\mu^2}
\left[\ln\left(\frac{Q^2}{\mu^2}\right)A+B+D_1\ln\frac{Q^2}{P_T^2R_1^2}+
D_2\ln\frac{Q^2}{P_T^2R_2^2}\right]\ , \label{su}
\end{eqnarray}
where $R_{1,2}$ are the jet radii of the two jets, respectively. In practice the jets are of course reconstructed with the same radius $R$ but to clarify the structure of our calculation we use two different radii $R_{1,2}$ to differentiate between the dijets. At one-loop order, $A=C_A \frac{\alpha_s}{\pi}$, $B=-2C_A\beta_0\frac{\alpha_s}{\pi}$ for a gluon-gluon initial state, $A=C_F \frac{\alpha_s}{\pi}$, $B=\frac{-3C_F}{2}\frac{\alpha_s}{\pi}$ for a quark-quark initial state, and $A=\frac{(C_F+C_A) }{2}\frac{\alpha_s}{\pi}$, $B=(\frac{-3C_F}{4}-C_A\beta_0)\frac{\alpha_s}{\pi}$ for a gluon-quark initial state. In addition, $D_{1,2}=C_A\frac{\alpha_s}{2\pi}$ for a gluon jet and $D_{1,2}=C_F\frac{\alpha_s}{2\pi}$ for a quark jet, respectively. Here, $\beta_0=(11-2N_f/3)/12$, with $N_f$ being the number of effective light quarks.

The resummation formula in Eq.~(\ref{resum}) was obtained in Refs.~\cite{Sun:2014gfa,Sun:2015doa} by a detailed analysis of the soft gluon radiation at one-loop order. The leading contributions from soft gluon radiation can be factorized into the associated TMD parton distributions and can be resummed by solving the relevant evolution equations. At NLL, the soft gluon radiation is factorized into the soft factor ${\bf S}$ which is given by a matrix in the color space of the partonic channels. The matrix form of the factorization is the same as was found for threshold resummation for the dijet production in proton-proton collisions~\cite{Kidonakis:1998bk,Kidonakis:1998nf,Kidonakis:2000gi,Kelley:2010fn,Catani:2013vaa,Hinderer:2014qta}. 

It is known that TMD factorization in dijet production in hadronic collisions is highly nontrivial and that there are potential factorization breaking effects~\cite{Boer:2003tx,Qiu:2007ey,Collins:2007nk,Rogers:2010dm,Bacchetta:2005rm,Vogelsang:2007jk,Bomhof:2007su,Catani:2011st,Mitov:2012gt,Schwartz:2017nmr,Schwartz:2018obd}. First, non-global logarithms (NGLs)~\cite{Dasgupta:2001sh,Dasgupta:2002bw} start to contribute to the cross section at two-loop order. It has been shown that they cannot easily be included into a factorization formula, although numerical simulations can be made and their contribution can be taken into account~\cite{Banfi:2003jj,Banfi:2008qs}. In addition, TMD factorization will be explicitly broken at three-loop order for the unpolarized cross section. This leads to a modification of the coefficient $A^{(3)}$ in the above Sudakov form factor~\cite{Collins:2007nk,Becher:2010tm,Catani:2011st,Mitov:2012gt,Rothstein:2016bsq,Schwartz:2017nmr,Schwartz:2018obd,Sun:2015doa}.

Factorization breaking effects are particularly evident for the single transverse spin asymmetry (SSA) in dijet production~\cite{Collins:2007nk}, $\Delta\sigma(S_\perp)=(\sigma(S_\perp)-\sigma(S_\perp))/2$, where $S_\perp$ represents the transverse polarization vector for one of the incoming nucleons. The SSA for this process is expressed as $\Delta\sigma(S_\perp)\propto \epsilon^{\alpha\beta} S_\perp^\alpha q_\perp^\beta$, i.e., the total transverse momentum of the two jets $q_\perp$ will have a preferred direction~\cite{Boer:2003tx}. This asymmetry is sensitive to the parton's Sivers function where the transverse momentum distribution is correlated with the transverse polarization vector~\cite{Sivers:1989cc}. In Refs.~\cite{Bacchetta:2005rm,Bomhof:2007su}, all initial/final state interaction contributions to the SSA were factorized into a complicated gauge link structure associated with the quark Sivers function for the polarized nucleon. However, for the double spin asymmetries involving two Sivers functions, it was shown explicitly that the generalized gauge-link approach to TMD factorization does not apply~\cite{Rogers:2010dm}. 

Ref.~\cite{Qiu:2007ey} provided an understanding of the SSA from the twist-three framework where the Qiu-Sterman-Efremov-Tereyav matrix elements are the basic ingredients~\cite{Efremov:1981sh,Efremov:1984ip,Qiu:1991pp,Qiu:1991wg, Qiu:1998ia}, and the high momentum Sivers function is generated by collinear gluon radiation. In particular, it was shown in Refs.~\cite{Qiu:2007ey,Efremov:1984ip} that the collinear gluon radiations parallel to the incoming hadrons can be factorized into the associated TMD parton distribution functions. It was also suggested that a factorization formula similar to that in the unpolarized case
may hold for the single spin dependent differential cross section~\cite{Qiu:2007ey},
\begin{eqnarray}
\frac{d\Delta\sigma(S_\perp)}
     {d\Omega d^2\vec{q}_\perp}
&=&
\frac{\epsilon^{\alpha\beta}S_\perp^\alpha q_\perp^\beta}
     {\vec{q}^2_\perp}
\sum\limits_{abcd}
\int d^2p_{1\perp}d^2p_{2\perp}d^2\lambda_\perp
\nonumber \\
&&\times
\frac{\vec{p}_{2\perp}\cdot \vec{q}_\perp}{M_P}\,
   x_2\, f_{1Tb}^{\perp(\rm SIDIS)}(x_2,p_{2\perp})\,
   x_1\, f_a^{(\rm SIDIS)}(x_1,p_{1\perp})
\label{e4}\\
&&\times
\left[S_{ab\to cd}(\lambda_\perp)\,
      H_{ab\to cd}^{\rm Sivers}(P_\perp^2)\right]_c\,
\delta^{(2)}(\vec{p}_{1\perp}+\vec{p}_{2\perp}+
\vec{\lambda}_\perp-\vec{q}_\perp) \, .
\nonumber
\end{eqnarray}
Here $f_{1Tb}^{\perp(\rm SIDIS)}$ and $f_a^{(\rm SIDIS)}$ denote the transverse-spin dependent TMD quark Sivers function and the unpolarized TMD parton distribution, respectively. These TMD parton distribution functions were chosen following their definitions in semi-inclusive deep-inelastic scattering (SIDIS) with future pointing gauge links. Although it was not explicitly shown in Ref.~\cite{Qiu:2007ey}, a matrix form of the factorization was suggested, where $H_{ab\to cd}^{\rm Sivers}$ and $S_{ab\to cd}$ are partonic hard and soft factors and the $[\quad ]_c$ represents a trace in color space between the hard and soft factors, similar to the unpolarized case in Eq.~(\ref{resum}). 

In order to check the factorization formula of Eq.~(\ref{e4}), it is important to carry out the calculation of soft gluon radiation. Soft gluon emissions contribute in a nontrivial way to the factorization formula. In particular, it will be crucial to show that these contributions can be included in the soft factor in the matrix form of the factorization formula. The goal of the current paper is to derive the soft gluon radiation contribution at one-loop order. As mentioned above, in Ref.~\cite{Qiu:2007ey} it was shown that collinear gluon radiation associated with the incoming nucleons can be treated following the general factorization arguments. This indicates that factorization holds in the leading logarithmic approximation (LLA). However, in order to obtain also all the subleading logarithmic contributions, we need to consider the soft gluon radiation as well. After including soft gluon radiation, we will obtain the complete double logarithmic result.

Our calculations presented in this work show that the factorization and resummation is expected to be valid at LLA. However, factorization breaking effects will emerge at NLL accuracy, in the sense that the contributions from soft gluon radiation cannot be factorized into the same soft factor as for the unpolarized case. This implies that beyond the LLA, we do not have a factorization formula for $\Delta \sigma$ as in Eq.~(\ref{resum}), at least not in the standard way with a spin independent soft factor.

In the LLA, we can express the spin dependent differential cross section in terms of the Fourier transform $b_\perp$ variable,
\begin{eqnarray}
\frac{d\Delta\sigma(S_\perp)}
     {d\Omega d^2\vec{q}_\perp}
&=&\epsilon^{\alpha\beta}S_\perp^\alpha\sum_{abcd}\int\frac{d^2\vec{b}_\perp}{(2\pi)^2}
e^{i\vec{q}_\perp\cdot
\vec{b}_\perp}W_{ab\to cd}^{T\beta}(x_1,x_2,b_\perp)\ .
\end{eqnarray}
Here, we neglect the $Y$-term contribution compared to the unpolarized case above. In this work we show that the leading logarithmic factorization of $W^{T\beta}$ takes the form
\begin{eqnarray}
W_{ab\to cd}^{T\beta}\left(x_1,x_2,b_\perp\right)|_{\rm LLA'}&=&\frac{ib_\perp^\beta}{2}x_1\,f_a(x_1,\mu=b_0/b_\perp)
x_2\, T_{Fb}(x_2,x_2,\mu=b_0/b_\perp)\nonumber\\
&&\times H_{ab\to cd}^{\rm Sivers} e^{-S_{\rm Sud}^T(Q^2,b_\perp)} \ ,\label{resumspin}
\end{eqnarray}
where $S_{\rm Sud}^T(Q^2,b_\perp)$ can be written in analogy to Eq.~(\ref{su}),
\begin{eqnarray}
S_{\rm Sud}^T(Q^2,b_\perp)=\int^{Q^2}_{b_0^2/b_\perp^2}\frac{d\mu^2}{\mu^2}
\left[\ln\left(\frac{Q^2}{\mu^2}\right)A+B+D_1\ln\frac{1}{R_1^2}+
D_2\ln\frac{1}{R_2^2}\right]\ . \label{sudakov-spin}
\end{eqnarray}
Here $A$, $B$, $D_{1,2}$ are the same as in Eq.~(\ref{su}). In the above equation, $T_F$ is the Qiu-Sterman matrix element which is also related to the transverse momentum-moment of the quark Sivers function. It is defined as follows
\begin{eqnarray}
T_F(x_2,x_2') &\equiv & \int\frac{d\zeta^-d\eta^-}{4\pi} e^{i(x_2
P_B^+\eta^-+(x_2'-x_2)P_B^+\zeta^-)}
\nonumber \\
&\times & \epsilon_\perp^{\beta\alpha}S_{\perp\beta} \,
\left\langle P_A,S|\overline\psi(0){\cal L}(0,\zeta^-)\gamma^+
\right. \label{TF}\\
&\times & \left. g{F_\alpha}^+ (\zeta^-) {\cal L}(\zeta^-,\eta^-)
\psi(\eta^-)|P_B,S\right\rangle  \ , \nonumber
\end{eqnarray}
where $F^{\mu\nu}$ represents the gluon field strength tensor. From the leading order derivation, we have  $\frac{1}{M_P}\int d^2k_{\perp}\, \vec{k}^2_\perp\, f_{1T}^{\perp(\rm SIDIS)}(x,k_\perp) = - T_F(x,x)$. For our soft gluon calculation at leading power, we take the correlation limit, i.e., we neglect all power corrections of the form $q_\perp/P_T$. In this limit, the leading double logarithm is proportional $1/q_\perp^2\times \ln(P_T^2/q_\perp^2)$. We will show that these contributions will be consistent with the resummation formula of Eq.~(\ref{resumspin}). Some sub-leading logarithmic terms can be factorized in this form as well. These include collinear gluon radiation associated with the incoming hadrons and the final state jets. The former can be resummed by including the scale evolution of the integrated parton distribution and the Qiu-Sterman matrix elements, e.g., by evaluating these distributions at the scale $\mu_b=b_0/b_\perp$. The latter are taken into account by the $\ln(1/R^2)$ terms in the Sudakov form factor $S_{\rm Sud}$ of Eq.~(\ref{su}). Therefore, Eq.~(\ref{resumspin}) is an improvement of the leading logarithmic approximation, to which we will refer as $\rm LLA'$ in the following.

The remainder of this paper is organized as follows. In Sec.~II, we will briefly review the soft gluon contribution to unpolarized dijet production. The basic formalism, including the Eikonal approximation, the phase space integrals to obtain the leading contributions, and the subtraction method to derive the soft gluon radiation associated with the final state jets will be introduced. Sec.~III contains the main new derivations of this work. We will carry out the calculation of the soft gluon radiation for the spin dependent differential cross sections. We introduce the general framework, the twist-three collinear expansion, and derive the soft gluon radiation amplitude in this formalism. We apply these techniques to different partonic channels and demonstrate that the leading double logarithmic contributions factorize, and we verify our resummation formula at LLA$'$. In Sec.~IV, we consider an example and show that factorization breaking effects appear at NLL, in particular, from soft gluon radiation that does not belong to the incoming hadrons and final state jets. In Sec.~V, we will present phenomenological results for the single spin asymmetries in dijet production at RHIC, and compare to recent STAR data~\cite{Abelev:2007ii,starpreliminary}. Finally, we will summarize our paper in Sec.~VI.

\section{Brief Review of Soft Gluon Radiation for the Unpolarized Case}

Dijet production at the leading order can be calculated from partonic $2\to 2$ processes,
\begin{equation}
a(p_1)+b(p_2)\to c(k_1)+d(k_2) \  ,
\end{equation}
where $p_{1,2}$ and $k_{1,2}$ are the momenta of the incoming and outgoing two partons, respectively. Their contributions to the cross section can be written as
\begin{eqnarray}
\frac{d^4\sigma}
{d\Omega d^2q_{\perp}}=\sum_{abcd}\sigma_0x_1\,f_a(x_1,\mu)
x_2\, f_b(x_2,\mu) {h}_{ab\to cd}^{(0)}\delta^{(2)}(q_\perp) \ ,
\end{eqnarray}
where the overall normalization of the differential cross section is $\sigma_0=\frac{\alpha_s^2\pi}{s^2}$. The partonic cross sections $h^{(0)}$ for all the production channels depend on the kinematic variables $\hat{s} = (p_1+p_2)^2$, $\hat{t} = (p_1-k_1)^2$ and $\hat{u} = (p_1-k_2)^2$. As mentioned above, at the leading order, they contribute to a delta function setting $q_\perp=0$, which corresponds to the back-to-back configuration of the two jets in the transverse plane. For soft gluon emissions, we can apply the leading power expansion and derive the dominant contribution from the Eikonal approximation~\cite{Mueller:2013wwa,Sun:2015doa}. For example, for the outgoing quark, antiquark and gluon lines, we obtain the following factors in the Eikonal approximation:
\begin{equation}
\frac{2k_i^\mu}{2k_i\cdot k_g+i\epsilon} g\ ,~~-\frac{2k_i^\mu}{2k_i\cdot k_g+i\epsilon}g \ ,~~\frac{2k_i^\mu}{2k_i\cdot k_g+i\epsilon}g \ ,
\end{equation}
respectively,  at one-loop order. Here $g$ is the strong coupling and the $k_i$ represent the momenta of the outgoing particles. For incoming quark, antiquark and gluon lines, we have,
\begin{equation}
-\frac{2p_1^\mu}{2p_1\cdot k_g-i\epsilon} g\ ,~~\frac{2p_1^\mu}{2p_1\cdot k_g-i\epsilon}g \
,~~\frac{2p_1^\mu}{2p_1\cdot k_g-i\epsilon} g \ ,
\end{equation}
respectively, where $p_1$ corresponds to the momentum of the incoming particle.

\begin{figure}[tbp]
\centering
\includegraphics[width=12cm]{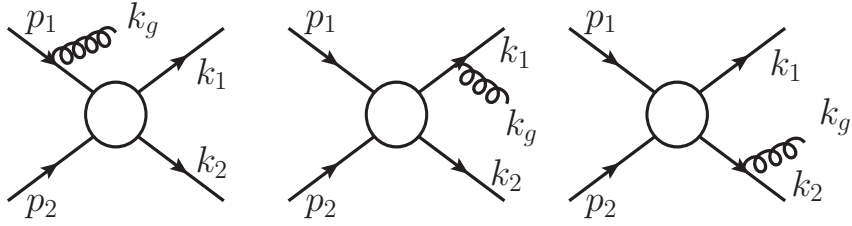}
\caption{Soft gluon radiation contributions to the finite imbalance transverse
momentum $q_\perp$: (a) initial state radiation, and (b), (c) final state radiation. Since we have chosen the gluon polarization vector along $p_2$, there
is no gluon radiation from the line with momentum $p_2$.}
\label{unpolarized}
\end{figure}

Following Ref.~\cite{Mueller:2013wwa,Sun:2015doa}, we choose physical polarizations of the soft gluon along the incoming particle with momentum $p_2$, so that the polarization tensor of the radiated gluon takes the following form:
\begin{equation}
\Gamma^{\mu\nu}(k_g)=\left(-g^{\mu\nu}+\frac{k_g^\mu p_2^\nu+k_g^\nu p_2^\mu}{k_g\cdot p_2}\right)  \ .\label{gluonpolarization}
\end{equation}
This choice will simplify the derivation since there is no soft gluon radiation from the incoming parton line $p_2$. Therefore, as shown in Fig.~\ref{unpolarized}, the leading contributions come from the initial state radiation from the line with momentum $p_1$, and the final state emissions from the lines $k_1$ and $k_2$. The contributions by these diagrams can be evaluated by taking the amplitudes squared of the Eikonal vertex with the polarization vector of the radiated gluon contracted with the above tensor. This leads to the following expressions for the soft gluon radiation contributions,
\begin{eqnarray}
\frac{2p_1^\mu}{2p_1\cdot k_g}\frac{2p_1^\nu}{2p_1\cdot k_g}\Gamma_{\mu\nu}
&=&S_g(p_1,p_2)\ ,\\
\frac{2k_1^\mu}{2k_1\cdot k_g}\frac{2k_1^\nu}{2k_1\cdot k_g}\Gamma_{\mu\nu}
&=&S_g(k_1,p_2)\ ,\\
\frac{2k_2^\mu}{2k_2\cdot k_g}\frac{2k_2^\nu}{2k_2\cdot k_g}
\Gamma_{\mu\nu} &=&S_g(k_2,p_2)\ , \\
2\frac{2k_1^\mu}{2k_1\cdot k_g}\frac{2p_1^\nu}{2p_1\cdot k_g}\Gamma_{\mu\nu}
&=&S_g(k_1, p_2)+S_g(p_1, p_2)-S_g(k_1, p_1)\ ,\\
2\frac{2k_2^\mu}{2k_2\cdot k_g}\frac{2p_1^\nu}{2p_1\cdot k_g}\Gamma_{\mu\nu}
&=&S_g(k_2, p_2)+S_g(p_1, p_2)-S_g(k_2, p_1)\ ,\\
2\frac{2k_1^\mu}{2k_1\cdot k_g}\frac{2k_2^\nu}{2k_2\cdot k_g}\Gamma_{\mu\nu}
&=&S_g(k_1, p_2)+S_g(k_2, p_2)-S_g(k_1, k_2)\ .
\end{eqnarray}
Here $S_g(p,q)$ is a short-hand notation for
\begin{equation}
S_g(p,q)=\frac{2p\cdot q}{p\cdot k_gq\cdot k_g} \ .
\end{equation}
When we integrate out the phase space of the radiated gluon to obtain the finite transverse momentum for the dijet imbalance, we have to exclude the contributions that belong to the jets. Therefore, only gluon radiation outside of the jets with radius $R_{1,2}$ contributes. These diagrams have been calculated in Refs.~\cite{Sun:2015doa}, where an offshellness was considered to regulate the collinear divergence associated with the jet within the narrow jet approximation~\cite{Aversa:1988vb,Mukherjee:2012uz}. In Ref.~\cite{Liu:2018trl,Liu:2020dct}, a subtraction method was employed to derive the soft gluon radiation contribution. For completeness, we show details of the derivation in the Appendix. Here, we list the final results:
\begin{eqnarray}
S_g(p_1,p_2)&\Rightarrow &\frac{\alpha_s}{2\pi^2}\frac{1}{q_\perp^2}\left(2\ln\frac{Q^2}{q_\perp^2}\right)\ ,\\
S_g(k_1,p_1)&\Rightarrow &\frac{\alpha_s}{2\pi^2}\frac{1}{q_\perp^2}\left[\ln\frac{Q^2}{q_\perp^2}+\ln\frac{1}{R_1^2}+\ln\left(\frac{\hat{t}}{\hat{u}}\right)
+\epsilon\left(\frac{1}{2}\ln^2\frac{1}{R_1^2}\right)\right]\ ,\label{e11}\\
S_g(k_2,p_1)&\Rightarrow &\frac{\alpha_s}{2\pi^2}\frac{1}{q_\perp^2}\left[\ln\frac{Q^2}{q_\perp^2}+\ln\frac{1}{R_2^2}+\ln\left(\frac{\hat{u}}{\hat{t}}\right)
+\epsilon\left(\frac{1}{2}\ln^2\frac{1}{R_2^2}\right)\right]\ ,\label{e21}\\
S_g(k_1,p_2)&\Rightarrow &\frac{\alpha_s}{2\pi^2}\frac{1}{q_\perp^2}\left[\ln\frac{Q^2}{q_\perp^2}+\ln\frac{1}{R_1^2}+\ln\left(\frac{\hat{u}}{\hat{}t}\right)
+\epsilon\left(\frac{1}{2}\ln^2\frac{1}{R_1^2}\right)\right]\ ,\label{e12}\\
S_g(k_2,p_2)&\Rightarrow &\frac{\alpha_s}{2\pi^2}\frac{1}{q_\perp^2}\left[\ln\frac{Q^2}{q_\perp^2}+\ln\frac{1}{R_2^2}+\ln\left(\frac{\hat{t}}{\hat{u}}\right)
+\epsilon\left(\frac{1}{2}\ln^2\frac{1}{R_2^2}\right)\right]\ ,\label{e22}\\
S_g(k_1,k_2)&\Rightarrow &\frac{\alpha_s}{2\pi^2}\frac{1}{q_\perp^2}\left[\ln\frac{1}{R_1^2}+\ln\frac{1}{R_2^2}+2\ln\left(\frac{\hat{s}^2}{\hat{t}\hat{u}}\right)
+\epsilon\left(\frac{1}{2}\ln^2\frac{1}{R_1^2}+\frac{1}{2}\ln^2\frac{1}{R_1^2}\right.\right.\nonumber\\
&&\left.\left.-4\ln\frac{\hat{s}}{-\hat{t}}\ln\frac{\hat{s}}{-\hat{u}}\right)\right]\ .\label{e22p}
\end{eqnarray}
Compared to the results in Ref.~\cite{Sun:2015doa}, the above results differ by a term proportional to $\epsilon\, \pi^2/6$. This is a result of the approximation made in Ref.~\cite{Sun:2015doa}.

\section{Soft Gluon Radiation for SSA: Leading Logarithmic Contributions}

In this section, we will investigate the soft gluon radiation contribution to the SSA in dijet production. The leading order analysis and collinear gluon radiation contributions have been studied in Ref.~\cite{Qiu:2007ey}. In the following, we will first review the leading order results and then derive the soft gluon radiation contribution.

\subsection{Leading Order Results}

\begin{figure}[t]
\begin{center}
\includegraphics[width=11cm]{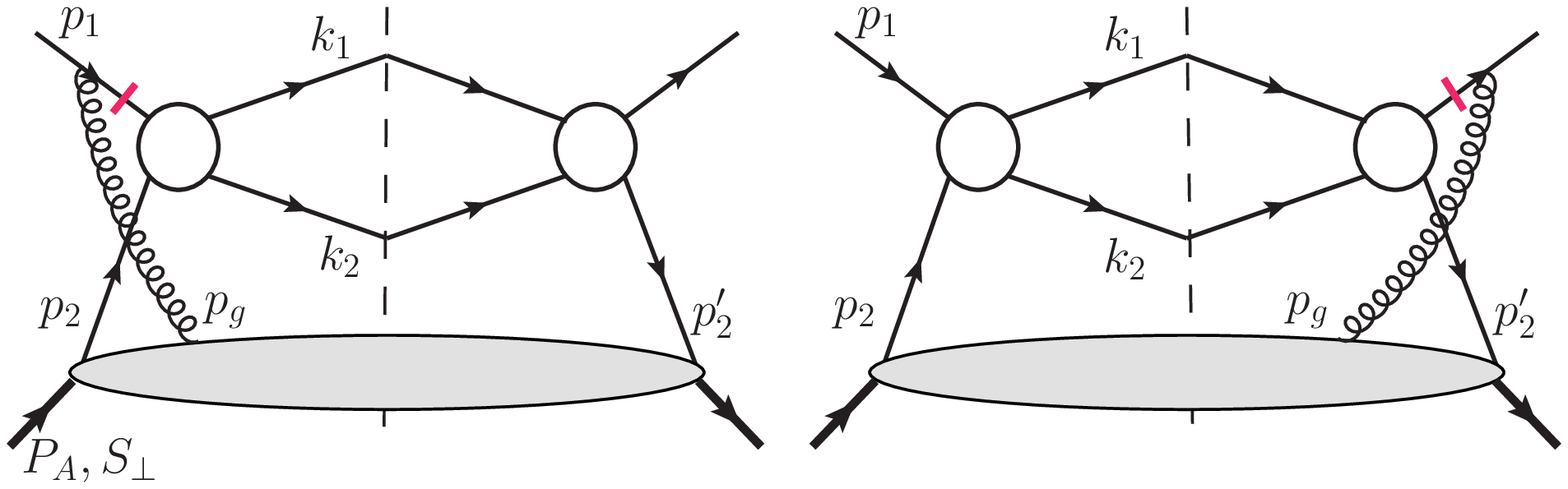}
\includegraphics[width=11cm]{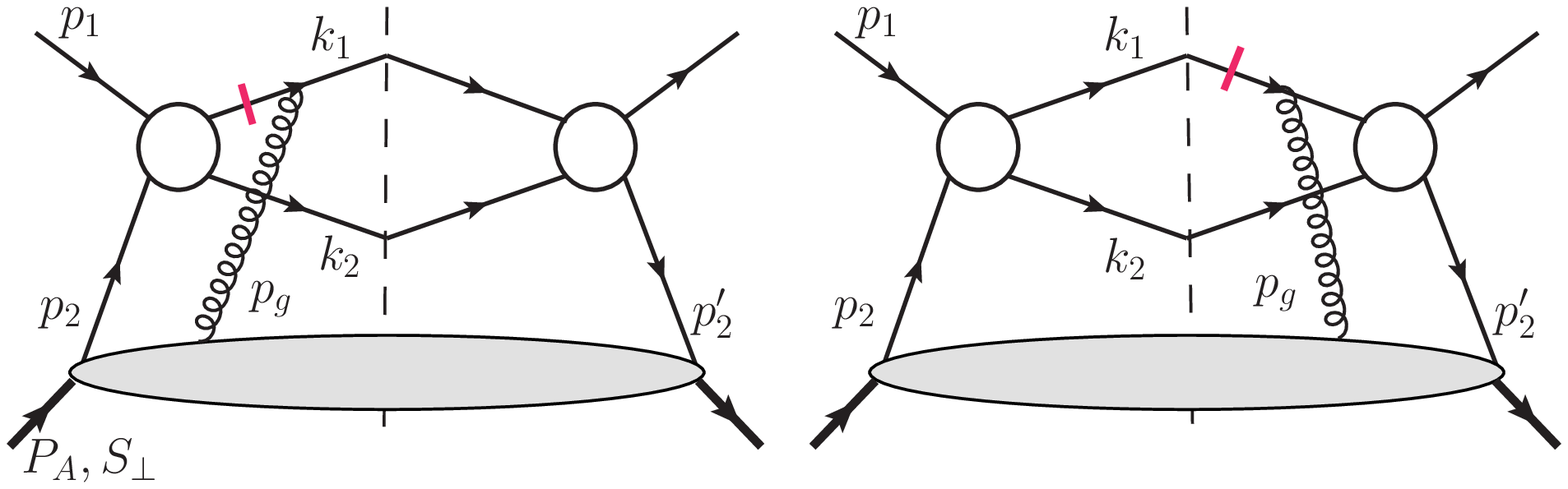}
\includegraphics[width=11cm]{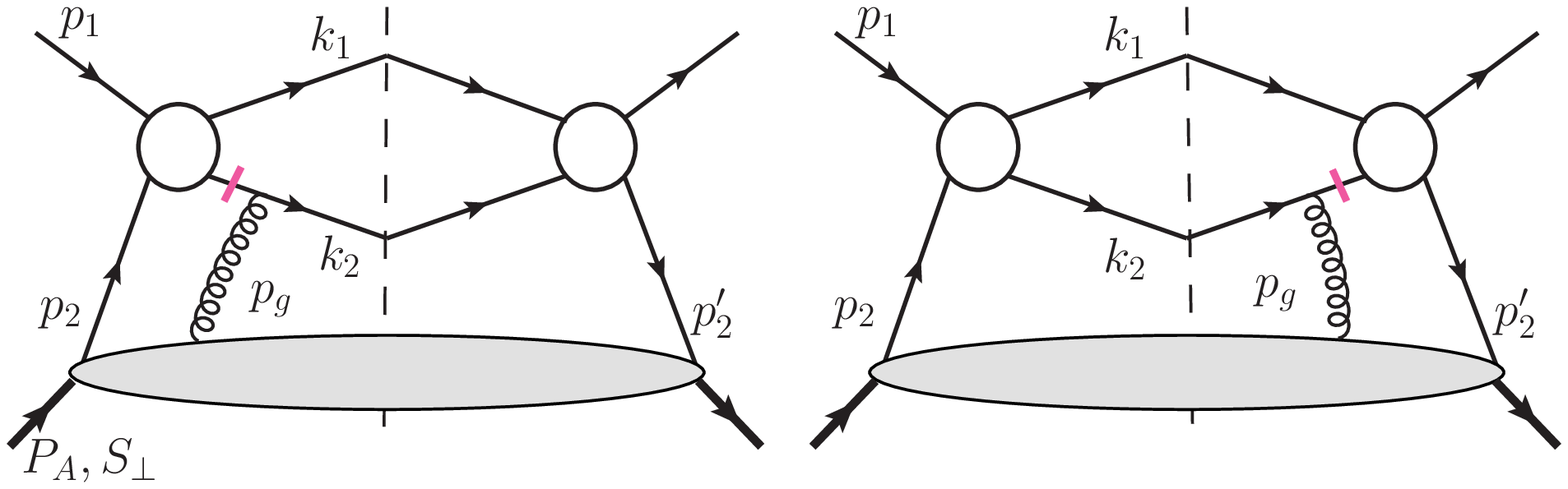}
\end{center}
\vskip -0.4cm \caption{\it Leading order diagrams for the initial and final state contributions to the SSA in dijet production. The red bars in these diagrams indicate the propagators that produce the necessary phase for the SSA. }
\label{leadingorder}
\end{figure}

The leading order results of Ref.~\cite{Qiu:2007ey} can be transformed into the factorization formula of Eq.~(\ref{resumspin}). For convenience, we show the diagrams that contribute to the SSA from initial and final state interaction effects in Figs.~\ref{leadingorder}. As demonstrated here, the SSA phases only come from the gluon attachments to the initial/final state partons. The leading order results derived in Ref.~\cite{Qiu:2007ey} can be obtained from Eq.~(\ref{e4}) by setting $\left[S_{ab\to cd}(\lambda_\perp)H_{ab\to cd}^{\rm Sivers}(P_T^2)\right]_c\equiv H_{ab\to cd}^{\rm Sivers}(P_T^2)$. After taking the Fourier transform to impact parameter $b_\perp$-space, we find the following leading order result:
\begin{eqnarray}
W_{ab\to cd}^{T\beta (0)}(b_\perp)=\frac{i b_\perp^\beta}{2} x_1f_a(x_1) x_2T_{Fb}(x_2,x_2) H_{ab\to cd}^{\rm Sivers} \ .
\end{eqnarray}
The hard part is written as
\begin{eqnarray}
H_{ab\to cd}^{\rm Sivers}&=&\frac{\alpha_s^2\pi}{\hat s^2}\sum_i
(C_I^i+C_{F1}^i+C_{F2}^i) h^i_{ab\to cd} \ , \label{e301}
\end{eqnarray}
where $i$ labels the different contributions to the hard factors $h^i$ by the various Feynman diagrams. Here the factors $C_I^i$ are for the initial state interaction for the single-spin dependent cross section, and $C_{F1}^i$ and $C_{F2}^i$ are for the final state interactions when gluon is attached to the lines with momentum $k_1$ and $k_2$, respectively. The explicit expressions for $C_I^i$, $C_{F1}^i$, $C_{F2}^i$ and $h^i$ are given in Ref.~\cite{Qiu:2007ey}.

Within the twist-three framework, we can also derive the hard factors at leading order by following a similar analysis as in Ref.~\cite{Qiu:2007ey}. The method for calculating the single transverse-spin asymmetry for hard scattering processes in the twist-three approach has been developed in Refs.~\cite{Qiu:1991pp,Qiu:1991wg, Qiu:1998ia,Ji:2006ub,Ji:2006vf,Ji:2006br,Kouvaris:2006zy,Eguchi:2006qz,Eguchi:2006mc,Koike:2006qv,Koike:2007rq,Koike:2007dg,Braun:2009mi,Kang:2008ey,Vogelsang:2009pj,Zhou:2008mz,Schafer:2012ra,Kang:2011mr,Sun:2013hua,Scimemi:2019gge}. The collinear expansion is the central step to obtain the final results. We perform our calculations in a covariant gauge. The additional gluon from the polarized hadron is associated with a gauge potential $A^\mu$, and one of the leading contributions comes from its component $A^+$. Thus, the gluon will carry longitudinal polarization. The gluon's momentum is dominated by $x_gP_A+p_{g\perp}$, where $x_g$ is the longitudinal momentum fraction with respect to the polarized proton. The contribution to the single-transverse-spin asymmetry arises from terms linear in $p_{g\perp}$ in the expansion of the partonic scattering amplitudes. When combined with $A^+$, these linear terms will yield $\partial^\perp A^+$, a part of the gauge field strength tensor $F^{\perp +}$ in Eq.~(\ref{TF}). Since $p_{g\perp}=p_{2\perp}'-p_{2\perp}$, the $p_{g\perp}$ expansion of the scattering amplitudes can be performed in terms of the transverse momenta $p_{2\perp}$ and $p_{2\perp}'$, which we can parametrize in the following way,
\begin{equation}
p_{2}=x_2P_A+p_{2\perp},~~~p_{2}'=x_2'P_A+p_{2\perp}' \ . \label{e55}
\end{equation}
The leading order diagrams shown in Fig.~\ref{leadingorder} can be calculated following the general procedure discussed above. The method is similar to the analysis of Drell-Yan lepton pair production in Ref.~\cite{Kang:2011mr}. For example, to perform the Fourier transform of the SSA from transverse momentum to the impact parameter $b_\perp$-space, we take the total transverse momentum of the two jets $q_\perp$. In the leading order diagrams as shown in Fig.~\ref{leadingorder}, the total transverse momentum $q_\perp$ can be easily identified: $q_\perp=p_{2\perp}^{~\prime}$ for the left diagrams and $q_\perp=p_{2\perp}$ for the right diagrams. Because the phases are opposite to each other for the left and right diagram, their total contribution will lead to the expression $\epsilon_{\alpha\beta}S_\perp^\alpha q_\perp^\beta=\epsilon_{\alpha\beta}S_\perp^\alpha p_{g\perp}^\beta$. Using the definition of the twist-three matrix element, we find that the SSA contribution in the impact parameter $b_\perp$-space can be written as $i\epsilon_{\alpha\beta}S_\perp^\alpha b_\perp^\beta T_F(x_2,x_2)$. The hard factors can be calculated accordingly.

\subsection{Soft Gluon Radiation}

At one-loop order, we have to consider real gluon radiation associated with the production of the dijets. When the radiated gluons are parallel to the incoming partons' momenta, their contributions can be factorized into the associated parton distribution functions (from the unpolarized nucleon) or the polarized quark Sivers function (from the polarized nucleon). The gluon radiation will generate finite transverse momentum. According to the analysis of Ref.~\cite{Qiu:2007ey}, we can write down the spin-dependent differential cross section as
\begin{eqnarray}
\frac{d\Delta\sigma(S_\perp)}{d\Omega d^2 q_\perp}&=&-\sum_{abcd} H_{ab\to cd}^{\rm Sivers}\epsilon^{\alpha\beta}S_\perp^\alpha \frac{\alpha_s}{2\pi^2}\frac{q_\perp^\beta}{(q_\perp^2)^2}x_1x_2\nonumber\\
&&\times \left\{
f_a(x_1)\widetilde{\cal P}_{b'g\to bg}^{T(<)}\otimes T_{Fb'}(x_2,x_2)
+T_{Fb}(x_2,x_2)\widetilde{\cal P}_{a'\to a}^{(<)}\otimes f_{a'}(x_1)\right\} \ ,\label{e30}
\end{eqnarray}
where ${\widetilde{\cal P}}^{(<)}$ represents the collinear splitting kernel excluding the end point contribution. For the twist-three function it is given by
\begin{eqnarray}
\widetilde{\cal P}_{b'g\to bg}^{T(<)}\otimes T_{Fb'}(x_2,x_2)&=&\int \frac{dx}{x}
     \left\{\frac{1}{2N_c}\left[(1+\xi^2)\left(x\frac{\partial}{\partial x}T_{Fb'}(x,x)\right)+T_{Fb'}(x,x)\frac{2\xi^3-3\xi^2-1}{1-\xi}\right]
\nonumber\right.\\
&&\left. +\left(\frac{1}{2N_c}+C_F\right) T_{Fb'}(x,x-\hat x_g)
 \frac{1+\xi}{1-\xi}\right\}\ ,
\end{eqnarray}
where $\xi=x_2/x$ and $\hat x_g=(1-\xi)x$. A similar (albeit slightly simpler) expression holds for ${\cal P}_{a'\to a}^{(<)}\otimes f_{a'}(x_1)$. In Eq.~(\ref{e30}), the first term in the bracket comes from the collinear gluon radiation associated with the polarized nucleon, whereas the second term is associated with the unpolarized nucleon. An explicit calculation of all the relevant diagrams was presented in Ref.~\cite{Qiu:2007ey} for one particular channel ($qq'\to qq'$), and factorization arguments were given for all other channels. Only the so-called soft- and hard-gluonic poles are considered in the SSA calculations. However, all other pole contributions and $\widetilde{T}_F$ contributions can be analyzed as well and similar results are expected. 

In the following, we will focus on soft gluon radiation. The factorization of these contributions is more involved for several reasons. First, they contain double logarithms. In terms of transverse momentum distributions, we will find terms of the form  $1/q_\perp^2\ln(P_T^2/q_\perp^2)$. These double logarithmic terms come from gluon radiation associated with all external particles. The collinear factorization arguments in Ref.~\cite{Qiu:2007ey} do not apply to these soft gluon emissions. Second, we have to deal with the soft gluon radiation associated with the final state jets. Using recent developments for the unpolarized case, we will be able to derive their contributions to the spin-dependent cross sections. We will first discuss several general features of twist-three calculations of the soft gluon radiation contributions to the SSA, and then we will apply these to the different partonic channels.

\subsubsection{Generic Features of Twist-three Calculations}

\begin{figure}[t]
\begin{center}
\includegraphics[width=11cm]{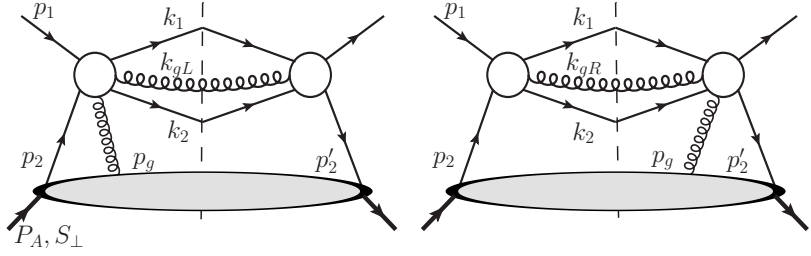}
\end{center}
\vskip -0.4cm \caption{\it Generic diagrams for the quark-quark scattering contributions to
the single transverse-spin dependent cross section within collinear factorization.}
\label{generic_diagrams}
\end{figure}

In Fig.~\ref{generic_diagrams}, we show the generic diagrams that need to be calculated to obtain the soft gluon real radiation contributions. We follow the same strategy as in Ref.~\cite{Sun:2015doa} to evaluate these diagrams. The radiated gluon carries transverse momentum $k_{g\perp}$ which will contribute to the total transverse momentum of the two jets. The spin-dependent differential cross section can for a given partonic channel be schematically written as
\begin{eqnarray}
\frac{d\Delta\sigma(S_\perp)}{d\Omega d^2 q_\perp}\Big|_{\textrm{soft}}^{(1)}={\epsilon^{\alpha\beta}S_\perp^\alpha}x_1f_a(x_1)x_2T_{Fb}(x_2,x_2) {\cal H}_{\textrm{twist-3}}^\beta (q_\perp,P_T;R)\ ,
\end{eqnarray}
for the one-loop soft gluon radiation, where $x_{1,2}$ are defined as for leading order kinematics. This is because the soft gluons do not modify the longitudinal momentum fractions of the incoming partons. The partonic cross section ${\cal H}_{\textrm{twist-3}}^\beta$ depends on the total transverse momentum $q_\perp$, the hard momentum scale represented by $P_T$ (or in general, $\hat s$, $\hat t$ and $\hat u$) and the jet radius $R$. Similar to the unpolarized case discussed in the last section, we have to exclude the gluon emission contributions belonging to the final state jets. Therefore, the partonic cross sections will depend on the jet size $R$. 

From the diagrams in Fig.~\ref{generic_diagrams}, we find that at this order the total transverse momentum of the two jets is equal and opposite to the transverse momentum of the radiated gluon: $\vec{q}_\perp=-\vec{k}_{g\perp}$. Therefore, finite $q_\perp$ also implies finite $k_{g\perp}$. The main objective of the following calculations is to obtain ${\cal H}_{\textrm{twist-3}}^\beta$ in the twist-three framework. Again, as briefly discussed above, we need to perform the collinear expansion for the incoming quark lines associated with the polarized nucleon. In the twist expansion, we take the limit of $k_{g\perp}\gg p_{2\perp}\sim p_{2\perp}^{~\prime}\sim p_{g\perp}$. Meanwhile, we are also working in the correlation limit of $q_\perp\sim k_{g\perp}\ll P_T$.  Therefore, the dominant contribution to the SSA comes from the expansion in powers of $p_{g\perp}/k_{g\perp}$. Any terms of the form $p_{g\perp}/P_T$ will be power suppressed in the correlation limit of $q_\perp\ll P_T$. 

To obtain the SSA for this process, the longitudinal gluon $p_g$ from the polarized nucleon needs to couple to the partonic scattering part to generate the necessary phase for a non-zero single spin asymmetry. Because we work at leading power in the limit of $q_\perp\ll P_T$, we can classify the gluon attachments into two types. First, the $p_g$ gluon attaches to one of the initial/final state partons which does not radiate the soft gluon $k_g$. Second, the $p_g$ gluon attachment and the soft gluon $k_g$ radiation happen on the same initial/final state parton. Here we discuss both cases, where the first type is easier to calculate, whereas the second type is somewhat more involved.

\begin{figure}[t]
\begin{center}
\includegraphics[width=11cm]{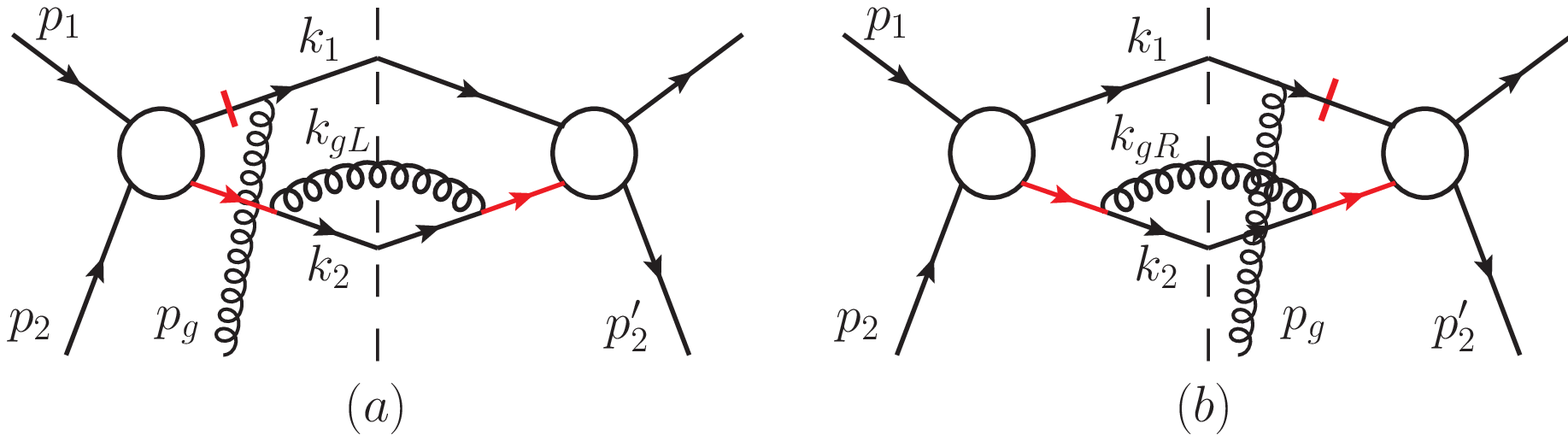}
\vskip 0.4cm
\includegraphics[width=11cm]{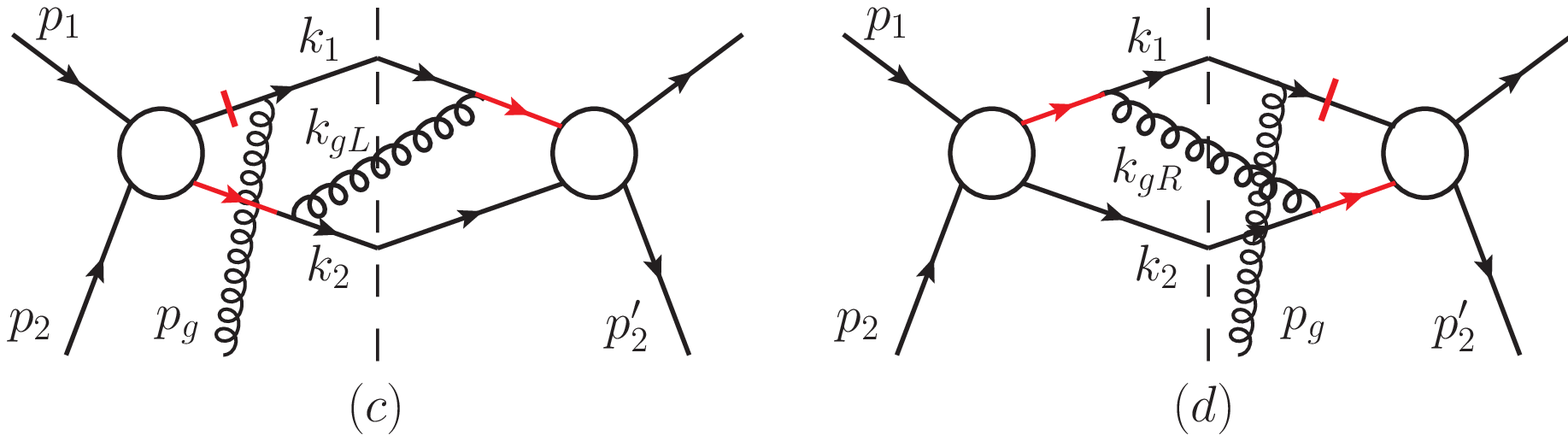}
\end{center}
\vskip -0.4cm \caption{\it Soft-gluonic pole contribution associated with the final state particle $k_1$ and the gluon radiation from the final state particle $k_2$. The dashed line in the middle indicates the final state particles which are on mass-shell. The left diagrams represent the contribution when the gluon is attached to the left side of the cut-line, whereas the right diagrams correspond to the attachment to the right side of the cut-line.}
\label{soft-pole-example}
\end{figure}

We first study the first type of diagrams. Particular examples are shown in Fig.~\ref{soft-pole-example}, where the pole contribution comes from the gluon attachment to the final state parton $k_1$ and the soft gluon is radiated off the lines with momentum $k_2$. Before an explicit evaluation, we would like to point out a number of important features which will help us to simplify the calculation. We will focus on the leading power contribution in the limit of small $q_\perp/P_T$. Therefore, the twist-three contributions only come from the $p_{i\perp}$-expansion associated with the radiated gluon line $k_g$. For example, the $p_{i\perp}$ dependence of the internal propagator (represented by the circle in Fig.~\ref{soft-pole-example}) will lead to a power suppressed contribution in the limit of $q_\perp\ll P_T$. Therefore, we only need to consider the $p_{i\perp}$-expansion in the propagators indicated by the red lines in Fig.~\ref{soft-pole-example}. In addition, the lines cut by red bars are the places where we pick up the pole contributions. The left and right diagrams give opposite contributions from these two poles, because they are on opposite sides of the cut-line. 

Because $k_1$ and $k_2$ are final sate observed momenta, it is convenient to keep them fixed in the $p_{i\perp}$-expansion. As a consequence, the momentum flow will go through the radiated gluon momentum. For convenience, we define $k_g$ as the radiated gluon momentum with $p_{i\perp}=0$, i.e., there is no $p_{i\perp}$ dependence in $k_g$. We label $k_{gL}$ as the momentum of the radiated gluon for the left diagram in Fig.~\ref{soft-pole-example}, and $k_{gR}$ for the right diagram, respectively. Due to the fact that the momentum flows are different for these two diagrams, $k_{gL}$ and $k_{gR}$ will be different as well. Each of them is constrained by the on-shell condition for the radiated gluon. For example, we know that $k_{gL\perp}=k_{g\perp}+p_{2\perp}'$. This gives the following momentum decomposition for $k_{gL}$:
\begin{eqnarray}
k_{gL}=k_g+\frac{\vec{k}_{g\perp}\cdot \vec{p}_{2\perp}^{~\prime}}{2p_2\cdot k_g}p_2+p_{2\perp}'\ ,
\end{eqnarray}
We find that $k_{gL}^2=0$ which satisfies the on-shell condition up to the linear term of $p_{i\perp}$. In the above expansion and the following calculations, we neglect all higher order terms of $p_{i\perp}$ beyond the linear terms. Similarly, we find for the right diagram of Fig.~\ref{soft-pole-example},
\begin{eqnarray}
k_{gR}=k_g+\frac{\vec{k}_{g\perp}\cdot \vec{p}_{2\perp}}{2p_2\cdot k_g}p_2+p_{2\perp}\ .
\end{eqnarray}
Once the kinematics are determined, we can proceed to calculate the soft gluon radiation contributions. This is similar to the unpolarized case. We multiply the Eikonal amplitudes of the diagrams shown in Fig.~\ref{soft-pole-example} and perform the collinear expansion of $p_{i\perp}$. For example, for the upper two diagrams, we have
\begin{eqnarray}
{\rm Left:}&&\frac{2k_{2\mu}}{(k_2+k_{gL})^2}\frac{2k_{2\nu}}{(k_2+k_{gL})^2}\Gamma^{\mu\nu}(k_{gL}) \ ,\\
{\rm Right:}&&\frac{2k_{2\mu}}{(k_2+k_{gR})^2}\frac{2k_{2\nu}}{(k_2+k_{gR})^2}\Gamma^{\mu\nu}(k_{gR}) \ ,
\end{eqnarray}
where $\Gamma^{\mu\nu}(k_g)$ was defined in Eq.~(\ref{gluonpolarization}). We stress that $k_{gL}$ and $k_{gR}$ depend on $p_{i\perp}$, which implies that $\Gamma^{\mu\nu}$ will as well. Because the left and right diagrams give contributions with opposite sign for the phase, which is necessary to generate the SSA for this process, we will add their $p_{i\perp}$ expansions with different signs. In the end, the total contribution from these two diagrams leads to the following expression:
\begin{eqnarray}
 {\rm Fig.~\ref{soft-pole-example}}(a,b):&&\frac{2k_{2\mu}}{(k_2+k_{gL})^2}\frac{2k_{2\nu}}{(k_2+k_{gL})^2}\Gamma^{\mu\nu}(k_{gL})-
\frac{2k_{2\mu}}{(k_2+k_{gR})^2}\frac{2k_{2\nu}}{(k_2+k_{gR})^2}\Gamma^{\mu\nu}(k_{gR})\nonumber\\
&&~~=-p_{g\perp}^\alpha\frac{2k_2\cdot p_2(k_{g\perp}^\alpha k_2\cdot p_2-k_{2\perp}^\alpha k_g\cdot p_2)}{(p_2\cdot k_g k_2\cdot k_g)^2}\nonumber\\
&&~~=-p_{g\perp}^\alpha 2(k_{g\perp}^\alpha-\xi_2k_{2\perp}^\alpha)\left(\frac{k_2\cdot p_2}{k_2\cdot k_g p_2\cdot k_g}\right)^2\ ,
\end{eqnarray}
where $\xi_2=k_g\cdot p_2/k_2\cdot p_2$. Noticing that $S_g(k_2,p_2)=4/(k_{g\perp}-\xi_2k_{2\perp})^2$, we can simplify the above result as
\begin{equation}
   {\rm Fig.~\ref{soft-pole-example}}(a,b):~~ p_{g\perp}^\alpha\frac{\partial}{\partial k_{g\perp}^\alpha}S_g(k_2,p_2) \ .
\end{equation}
Applying the twist-three procedure, the above leads to the following contribution to ${\cal H}_{\textrm{twist-3}}^\beta$,
\begin{equation}
    {\cal H}_{\textrm{twist-3}}^\beta|_{\rm Fig.~\ref{soft-pole-example}(a,b)}:~~\frac{\partial}{\partial k_{g\perp}^\beta}S_g(k_2,p_2)\ .
\end{equation}
To determine the leading power contribution from soft gluon radiation, we need to integrate out the phase space of the radiated gluon. We will come back to this point after completing the analysis of all relevant diagrams.

Now we turn to the lower two diagrams of Fig.~\ref{soft-pole-example}. Here, $k_{gL}$ and $k_{gR}$ are the same as in the upper two diagrams. However, the $p_{i\perp}$-expansion comes from different propagators,
\begin{eqnarray}
{\rm Left:}&&\frac{2k_{2\mu}}{(k_2+k_{gL})^2}\frac{2k_{1\nu}}{(k_1+k_{gL})^2}\Gamma^{\mu\nu}(k_{gL}) \ ,\\
{\rm Right:}&&\frac{2k_{1\mu}}{(k_1+k_{gR})^2}\frac{2k_{2\nu}}{(k_2+k_{gR})^2}\Gamma^{\mu\nu}(k_{gR}) \ .
\end{eqnarray}
We also find that the final result is a little more complicated than for the upper two diagrams. After some algebra, we find that it can be written as
\begin{eqnarray}
{\rm Fig.~\ref{soft-pole-example}}(c,d):~~ p_{g\perp}^\alpha\frac{\partial }{\partial k_{g\perp}^\alpha}\left[S_g(k_1,p_2)+S_g(k_2,p_2)-S_g(k_1,k_2)\right] \ .
\end{eqnarray}
The above derivations can be generalized to all other diagrams of the first type.

\begin{figure}[t]
\begin{center}
\includegraphics[width=11cm]{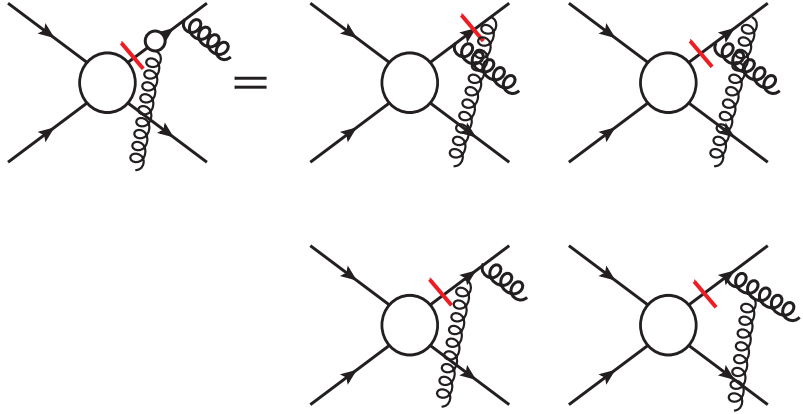}
\end{center}
\vskip -0.4cm \caption{\it Second type of diagrams associated with the final state interaction contribution with the jet $k_1$ (quark or gluon). The red bars represent the pole contributions. The first diagram gives the so-called soft-gluonic pole contributions, whereas the remaining three belong to the so-called hard gluonic pole contributions.}
\label{hard-pole-example0}
\end{figure}

For the second type of diagrams, the longitudinal gluon from the polarized proton can attach to both the final state jet and the radiated gluon. Therefore, we have both soft-gluonic pole and hard-gluonic pole contributions. One particular example is shown in Fig.~\ref{hard-pole-example0}. Here, the gluonic pole contributions come from the final state particle $k_1$ which also radiated a soft gluon to generate the leading power contribution in the correlation limit of small $q_\perp/P_T$. The first diagram corresponds to the soft-gluonic pole and the rest to the hard-gluonic pole. The soft gluon pole leads to a delta function $\delta(x_2-x_2')$. In general, the hard gluon pole leads to a different delta function. However, in the correlation limit, the hard pole reduces to the same delta function. For example, the hard pole (labeled by the red bar in Fig.~\ref{hard-pole-example0}) leads to the following kinematics:
\begin{eqnarray}
x_2'-x_2=\frac{k_g\cdot k_1}{P_A\cdot (k_1+k_g)} \ .
\end{eqnarray}
In the correlation limit~\footnote{This is also true in the collinear limit where the radiated gluon is parallel to the final state jet.}, i.e., $k_g^\mu\ll k_1^\mu$, the above reduces to $x_2'=x_2$. Therefore, the soft-gluonic and hard-gluonic pole contributions come from the same kinematics and there will be cancelations among them. These cancelations are very similar to those occurring for the SSA in the Drell-Yan process demonstrated in Ref.~\cite{Ji:2006vf}. In particular, because of the color factor $if_{abc}T^c=[T^a,T^b]$, we can decompose the last diagram into the other two diagrams associated with the hard-gluonic pole. The combination exactly cancels out the soft-gluonic pole contribution from the first diagram. We have explicitly checked this cancelation for all these diagrams. To carry out the calculation, we have to follow the transverse momentum flow, and perform the $p_{i\perp}$ expansion. The method is the same as that used to calculate the first type of diagrams: the kinematics of $k_g$ and $p_g$ will be determined from the on-shell conditions for $k_g$ and the pole contributions. More importantly, similar to the first type of diagrams, we only need to take into account the $p_{i\perp}$ expansion from the denominators of the relevant propagators, since the expansions of the numerators are power suppressed. Therefore, the calculations are the same for all partons in the initial/final state, regardless of whether they are quarks or gluons. Both have the same denominators. 

\begin{figure}[t]
\begin{center}
\includegraphics[width=11cm]{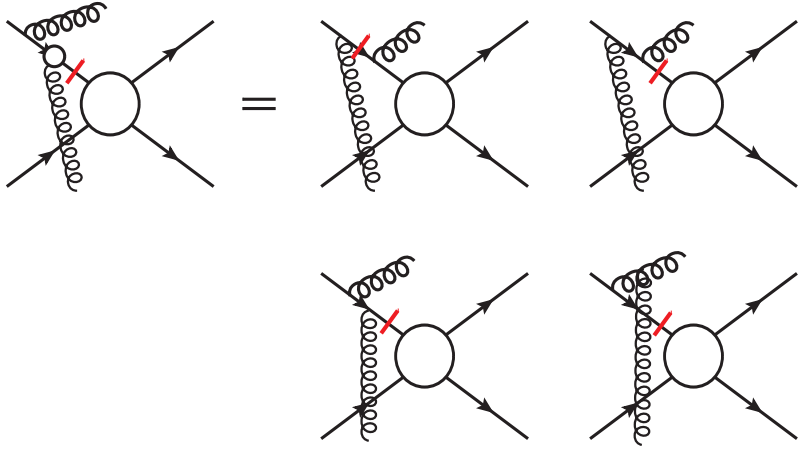}
\end{center}
\vskip -0.4cm \caption{\it Same as Fig.~\ref{hard-pole-example0} but for diagrams associated with the initial state interaction contributions with parton $p_1$ .}
\label{hard-pole-example1}
\end{figure}
\begin{figure}[t]
\begin{center}
\includegraphics[width=11cm]{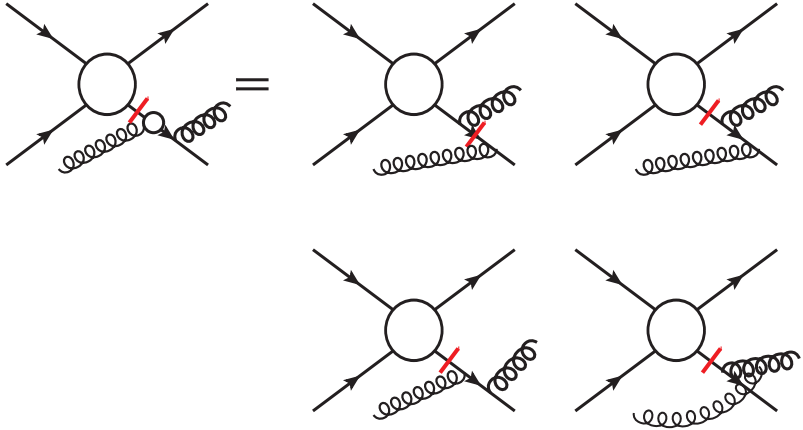}
\end{center}
\vskip -0.4cm \caption{\it Same as Fig.~\ref{hard-pole-example0} but for diagrams associated with the final state interaction contributions with $k_2$.}
\label{hard-pole-example2}
\end{figure}

In the end, the total contribution will be a combination of the last two diagrams with the color factor of the third one. In summary, we can represent all four diagrams as the one on the left side. This can be repeated for diagrams associated with the initial state interaction with $p_1$ and the final state interaction with $k_2$. We show the relevant diagrams in Figs.~\ref{hard-pole-example1} and \ref{hard-pole-example2}. These ``effective'' diagrams will be among the important ingredients for the final results.  

\begin{figure}[t]
\begin{center}
\includegraphics[width=11cm]{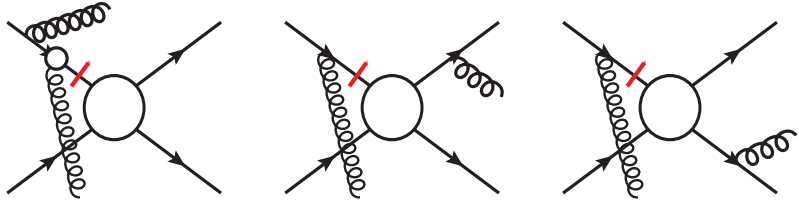}
\end{center}
\vskip -0.4cm \caption{\it Summary of gluon radiation diagrams for initial state interaction.}
\label{gluonradiation0}
\end{figure}

It is convenient to add the contributions from the first type and second type of diagrams. As an example, in Fig.~\ref{gluonradiation0}, we show those diagrams for the initial state interaction contributions. These diagrams show that we can add soft gluon radiation on top of the initial state interaction diagram. The core part is the same for these three, in particular, the associated color factors. We will work out the color factors for the different channels later on. Here, we focus on the kinematics of the soft gluon radiation contribution and in particular on the leading power contributions. 

As shown in Fig.~\ref{soft-pole-example}, the contributions to the SSA come from the interference between the diagrams of Fig.~\ref{gluonradiation0} and the diagrams of Fig.~\ref{unpolarized}. There will be diagrams where the longitudinal gluon is attached to the left side of the cut-line and to the right side of the cut-line. For convenience, we label these soft gluon radiation diagrams by their association with the external momenta. For example, we will label the first diagram of Fig.~\ref{gluonradiation0} by $p_1^\mu$, the second diagram by $k_1^\mu$ and the third by $k_2^\mu$. We label the diagrams in Fig.~\ref{unpolarized} similarly. The interference between the second diagram of Fig.~\ref{gluonradiation0} and the second 
one of Fig.~\ref{unpolarized} has been calculated above and the result is
\begin{eqnarray}
k_1^\mu k_1^\nu\Rightarrow \frac{\partial}{\partial k_{g\perp}^\beta}S_g(k_1,p_2) \ .
\end{eqnarray}
Similarly, we find the following result for the interference between the second diagram of Fig.~\ref{gluonradiation0} and the third diagram of Fig.~\ref{unpolarized}:
\begin{eqnarray}
k_1^\mu k_2^\nu\Rightarrow \frac{\partial}{\partial k_{g\perp}^\beta}\left[S_g(k_1, p_2)+S_g(k_2, p_2)-S_g(k_1, k_2)\right]\ .
\end{eqnarray}
The calculation for the interference between the first diagram of Fig.~\ref{gluonradiation0} and the diagrams in Fig.~\ref{unpolarized} is much more involved. However, after a lengthy derivation, we find the results are very similar to the above two, and they can be all summarized as follows:
\begin{eqnarray}
p_1^\mu p_1^\nu &\Rightarrow &
\frac{\partial}{\partial k_{g\perp}^\beta}S_g(p_1,p_2)\ ,\\
k_1^\mu k_1^\nu &\Rightarrow &
\frac{\partial}{\partial k_{g\perp}^\beta}S_g(k_1,p_2)\ ,\\
k_2^\mu k_2^\nu &\Rightarrow &
\frac{\partial}{\partial k_{g\perp}^\beta}S_g(k_2,p_2)\ ,\\
k_1^\mu p_1^\nu,p_1^\mu k_1^\nu &\Rightarrow &
\frac{\partial}{\partial k_{g\perp}^\beta}\left[S_g(k_1, p_2)+S_g(p_1, p_2)-S_g(k_1, p_1)\right]\ ,\\
k_2^\mu p_1^\nu,p_1^\mu k_2^\nu &\Rightarrow &
\frac{\partial}{\partial k_{g\perp}^\beta}\left[S_g(k_2, p_2)+S_g(p_1, p_2)-S_g(k_2, p_1)\right]\ ,\\
k_1^\mu k_2^\nu,k_2^\mu k_1^\nu &\Rightarrow &
\frac{\partial}{\partial k_{g\perp}^\beta}\left[S_g(k_1, p_2)+S_g(k_2, p_2)-S_g(k_1, k_2)\right]\ .
\end{eqnarray}
Interestingly, we note that there is a one-to-one correspondence between the above results and those for the gluon radiation contributions for the unpolarized case in Sec.~II. This is a very important feature to obtain the final factorization result for the SSA. 

\begin{figure}[t]
\begin{center}
\includegraphics[width=11cm]{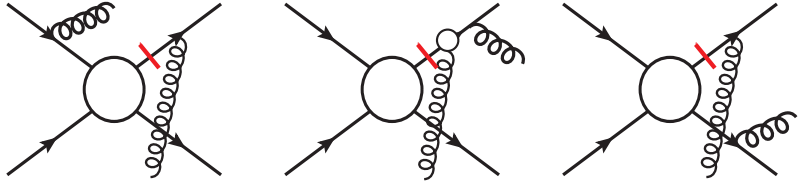}
\end{center}
\vskip -0.4cm \caption{\it Summary of the gluon radiation diagrams for the final state interaction contributions with $k_1$.}
\label{gluonradiation1}
\end{figure}
\begin{figure}[t]
\begin{center}
\includegraphics[width=11cm]{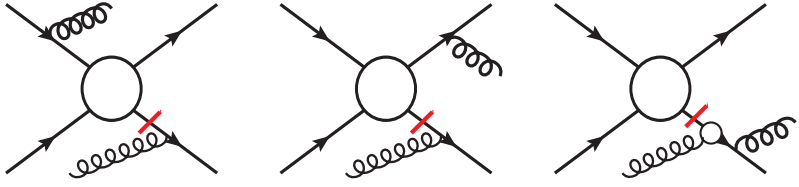}
\end{center}
\vskip -0.4cm \caption{\it Summary of the gluon radiation diagrams for the final state interaction contributions with $k_2$.}
\label{gluonradiation2}
\end{figure}

The above analysis can be extended to diagrams with final state interaction contributions with $k_1$ and $k_2$. For completeness, we show the diagrams in Fig.~\ref{gluonradiation1} and~\ref{gluonradiation2}. The final results are the same as those given above for the initial state interaction contributions. 

As mentioned above, in order to derive the leading power contribution to the SSA for this process from the above terms, we need to integrate over the phase space of the radiated gluon, where we keep the transverse momentum $\vec{q}_\perp=-\vec{k}_{g\perp}$. Let us first work out the simple example of the $p_1^\mu p_1^\nu$ term. This term is similar to that for Drell-Yan lepton pair production calculated in Ref.~\cite{Ji:2006vf}. The phase space integral takes the following form
\begin{eqnarray}
{\cal H}_{\textrm{twist-3}}^\beta\big|_{p_1^\mu p_1^\nu}&=&\frac{g^2}{2}\int \frac{d^3k_g}{(2\pi)^32E_{k_g}}\delta^{(2)} (q_\perp+k_{g\perp})\frac{\partial}{\partial k_{g\perp}^\beta}
S_g(p_1,p_2)\nonumber\\
&=&\frac{\alpha_s}{2\pi^2}\int_{\xi_0}^1\frac{d\xi}{\xi}\frac{\partial}{\partial k_{g\perp}^\beta}\frac{1}{\vec{k}_{g\perp}^2}|_{\vec{k}_{g\perp}=-\vec{q}_\perp} \ ,
\end{eqnarray}
where $\xi=k_g\cdot p_2/p_1\cdot p_2$ and the lower limit of the $\xi$-integral comes from the kinematic limit $\xi_0=k_{g\perp}^2/Q^2$. Working out the integral, we arrive at
\begin{eqnarray}
{\cal H}_{\textrm{twist-3}}^\beta\big|_{p_1^\mu p_1^\nu}&=&\frac{\alpha_s}{2\pi^2}\frac{2q_\perp^\beta}{(q_\perp^2)^2}\ln\frac{Q^2}{q_\perp^2} \ ,
\end{eqnarray}
which is consistent with the double logarithmic behavior for Drell-Yan lepton pair production calculated in Ref.~\cite{Ji:2006vf}. 

On the other hand, we can also carry out the above integral using integration by parts as follows:
\begin{eqnarray}
{\cal H}_{\textrm{twist-3}}^\beta\big|_{p_1^\mu p_1^\nu}
&=&\frac{\alpha_s}{2\pi^2}\left[\frac{\partial}{\partial k_{g\perp}^\beta}\int_{\xi_0}^1\frac{d\xi}{\xi}\frac{1}{\vec{k}_{g\perp}^2}+\frac{\partial \xi_0}{\partial k_{g\perp}^\beta}\frac{1}{\xi_0\vec{k}_{g\perp}^2}\right]_{\vec{k}_{g\perp}=-\vec{q}_\perp} \nonumber\\
&=&\frac{\alpha_s}{2\pi^2}\left[\frac{\partial}{\partial k_{g\perp}^\beta}\left(\frac{1}{k_{g\perp}^2}\ln\frac{Q^2}{k_{g\perp}^2}\right)+\frac{1}{\vec{k}_{g\perp}^2}\frac{\partial \ln k_{g\perp}^2 }{\partial k_{g\perp}^\beta}\right]_{\vec{k}_{g\perp}=-\vec{q}_\perp}\nonumber\\
&=&\frac{\alpha_s}{2\pi^2}\left[\frac{\partial}{\partial k_{g\perp}^\beta}\left(\frac{1}{k_{g\perp}^2}\right)\ln\frac{Q^2}{k_{g\perp}^2}\right]_{\vec{k}_{g\perp}=-\vec{q}_\perp}\ .
\end{eqnarray}
Of course, this gives the same result as above. However, this provides a convenient way to derive other terms as well. The rule is that the derivative only acts on the $1/k_{g\perp}^2$, not on the logarithmic terms. Taking the example of $S_g(k_1,p_2)$ and $S_g(k_1,p_1)$, we can follow the strategy and construct the following two terms:
\begin{eqnarray}
{\cal T}^\beta(k_1)&=&\frac{\alpha_s}{8\pi^2}\int\frac{d\xi}{\xi}\frac{\partial}{\partial k_{g\perp}^\beta}
\left[S_g(k_1,p_2)+S_g(k_1,p_1)\right]\ ,\\
{\cal R}^\beta(k_1)&=&\frac{\alpha_s}{8\pi^2}\int\frac{d\xi}{\xi}\frac{\partial}{\partial k_{g\perp}^\beta}
\left[S_g(k_1,p_2)-S_g(k_1,p_1)\right]\ .
\end{eqnarray}
In the above two equations, ${\cal R}^\beta$ does not contain a $\ln(1/k_{g\perp}^2)$ term. Therefore, we can perform the integration by parts directly and obtain the final result
\begin{eqnarray}
{\cal R}^\beta(k_1)&=&\frac{\alpha_s}{8\pi^2}\frac{\partial}{\partial k_{g\perp}^\beta}\int\frac{d\xi}{\xi}
\left[S_g(k_1,p_2)-S_g(k_1,p_1)\right]\nonumber\\
&=&\frac{\alpha_s}{2\pi^2}\frac{\partial}{\partial k_{g\perp}^\beta}\left(\frac{1}{k_{g\perp}^2}\ln\frac{\hat u}{\hat t}\right)\ .
\end{eqnarray}
For ${\cal T}^\beta$, we notice that $S_g(k_1,p_2)+S_g(k_1,p_1)=S_g(p_1,p_2)+4\vec{k}_{1\perp}\cdot \vec{k}_{g\perp}/(k_{g\perp}^2k_1\cdot k_g)$, where the first term has been calculated above and the second term does not have a term $\ln(1/k_{g\perp})$. Applying this, we arrive at the following result for ${\cal T}^\beta$:
\begin{eqnarray}
{\cal T}^\beta(k_1)&=&\frac{\alpha_s}{2\pi^2}\frac{\partial}{\partial k_{g\perp}^\beta}\left(\frac{1}{k_{g\perp}^2}\right)\left[\ln\frac{Q^2}{k_{g\perp}^2}+\ln\frac{1}{R_1^2}\right] \ .
\end{eqnarray}
Note that in practice we do the algebra and phase space integration in dimensional regularization in $d=4-2\epsilon$ dimensions. For simplicity, we are not displaying here terms of ${\cal O}(\epsilon)$. From the results for ${\cal T}^\beta$ and ${\cal R}^\beta$, we are able to derive the corresponding results for the terms associated with $S_g(k_1,p_1)$ and $S_g(k_1, p_2)$. Following the same technique, we can also derive the results for $S_g(k_2,p_2)$ and $S_g(k_2,p_1)$. For $S_g(k_1,k_2)$, since it does not contain a $\ln(1/k_{g\perp}^2)$ term, we can directly carry out the derivative with respect to $k_{g\perp}^\beta$. Finally, we summarize all results here:
\begin{eqnarray}
{\cal H}_{\textrm{twist-3}}^\beta\big|_{p_1^\mu p_1^\nu} &=&\frac{\alpha_s}{2\pi^2}\frac{q_\perp^\beta}{(q_\perp^2)^2}2\ln\frac{Q^2}{q_\perp^2} ,\label{p1p1}\\
{\cal H}_{\textrm{twist-3}}^\beta\big|_{k_1^\mu k_1^\nu} &=&\frac{\alpha_s}{2\pi^2}\frac{q_\perp^\beta}{(q_\perp^2)^2}\left[
\ln\frac{Q^2}{q_\perp^2}+\ln\frac{1}{R_1^2}+\ln\frac{\hat u}{\hat t}\right]\ ,\label{k1k1}\\
{\cal H}_{\textrm{twist-3}}^\beta\big|_{k_2^\mu k_2^\nu} &=&\frac{\alpha_s}{2\pi^2}\frac{q_\perp^\beta}{(q_\perp^2)^2}\left[ \ln\frac{Q^2}{q_\perp^2}+\ln\frac{1}{R_2^2}+\ln\frac{\hat t}{\hat u}\right]\ ,\label{k2k2}\\
{\cal H}_{\textrm{twist-3}}^\beta\big|_{k_1^\mu p_1^\nu,p_1^\mu k_1^\nu} &=&\frac{\alpha_s}{2\pi^2}\frac{q_\perp^\beta}{(q_\perp^2)^2}\left[
2\ln\frac{Q^2}{q_\perp^2}+2\ln\frac{\hat u}{\hat t}\right]\ ,\label{k1p1}\\
{\cal H}_{\textrm{twist-3}}^\beta\big|_{k_2^\mu p_1^\nu,p_1^\mu k_2^\nu} &=&\frac{\alpha_s}{2\pi^2}\frac{q_\perp^\beta}{(q_\perp^2)^2}\left[ 
2\ln\frac{Q^2}{q_\perp^2}+2\ln\frac{\hat t}{\hat u}\right]\ ,\label{p2p1}\\
{\cal H}_{\textrm{twist-3}}^\beta\big|_{k_1^\mu k_2^\nu,k_2^\mu k_1^\nu} &=&\frac{\alpha_s}{2\pi^2}\frac{q_\perp^\beta}{(q_\perp^2)^2}\left[ 
2\ln\frac{Q^2}{q_\perp^2}-2\ln\frac{\hat s^2}{\hat t\hat u}\right]\ . \label{k1k2}
\end{eqnarray}
Again, we have neglected the terms of ${\cal O}(\epsilon)$ for simplicity. It is interesting to note that all of the above terms contribute to the leading double logarithms, whereas only $k_1^\mu k_1^\nu$ and $k_2^\mu k_2^\nu$ contribute to the jet related logarithms. Therefore, we need to consider all of them to derive the leading double logarithmic contributions. Next, we need to combine the above results with the associated color factors for the different channels in order to obtain the contributions to the SSA.

\subsubsection{$q q'\to qq'$ channel}

Let us first derive the SSA for the simplest channel, $qq'\to qq'$, the quark-quark scattering with different flavors. This channel only has a $t$-channel diagram. The leading order results have been calculated in Ref.~\cite{Qiu:2007ey}. The hard factor is given by
\begin{eqnarray}
H_{qq'\to qq'}^{\rm Sivers}=\frac{\alpha_s^2\pi}{\hat s^2}\frac{N_c^2-5}{4N_c^2}\frac{2(\hat s^2+\hat u^2)}{\hat t^2} \ ,\label{hqq'}
\end{eqnarray}
where the color factors for the initial and final state interactions are: $C_I=-\frac{1}{2N_c^2}$, $C_{F1}=\frac{N_c^2-2}{4N_c^2}$, $C_{F2}=-\frac{1}{4N_c^2}$.

For the initial state interaction contributions, we calculate the interference between the diagrams in Fig.~\ref{unpolarized} and Fig.~\ref{gluonradiation0}. We obtain the following associated color factors:
\begin{eqnarray}
p_1^\mu p_1^\nu &\Rightarrow& C_F \ ,\\
k_1^\mu k_1^\nu &\Rightarrow& C_F \ ,\\
k_2^\mu k_2^\nu &\Rightarrow& C_F \ ,\\
p_1^\mu k_1^\nu &\Rightarrow& -\frac{1}{2N_c} \ ,\\
p_1^\mu k_2^\nu &\Rightarrow& \frac{3}{2}C_F-\frac{C_A}{2} \ ,\\
k_1^\mu k_2^\nu &\Rightarrow& \frac{3}{2}C_F-C_A \ .
\end{eqnarray}
In order to obtain the final results, we multiply the leading power contributions of Eqs.~(\ref{p1p1})-(\ref{k1k2}) with the associated color factors. Adding these results, we obtain the leading contribution from soft gluon radiation which can be written as
\begin{eqnarray}
{\cal H}_{\textrm{twist-3}}^{\beta(C_I)}({qq'\to qq'})&=&C_Ih_{q_iq_j\to q_iq_j}^{(0)}\frac{\alpha_s}{2\pi^2}\frac{-q_\perp^\beta}{(q_\perp^2)^2}\left[
2C_F\ln\frac{Q^2}{q_\perp^2}+C_F\left(\ln\frac{1}{R_1^2}+\ln\frac{1}{R_2^2}\right)\right]  \ . \label{qqsoft}
\end{eqnarray}
Here we only kept the terms relevant at $\rm LLA'$, and we have $h_{q_iq_j\to q_iq_j}^{(0)}=\frac{\alpha_s^2\pi}{\hat s^2}\frac{2(\hat s^2+\hat u^2)}{\hat t^2}$. We will come back to the remaining terms in Sec.~IV when we discuss factorization breaking effects. The minus sign in the above equation is due to the fact fact that $C_I$ was computed on the basis of the quark Sivers function for the SIDIS process, which has an opposite sign compared to the normalization of the twist-three matrix element $T_F$. It appears that the terms in Eq.~(\ref{qqsoft}) do have a clear factorization structure that includes the leading double logarithmic term and the terms associated with the final state jets. The latter are represented by logarithmic terms of the jet radii $R_{1,2}$.

\begin{table}[t]
\caption{{\it The color factors for the soft gluon radiation interference diagrams for the $q q'\to qq'$ channel. The different rows show the results for the unpolarized case as well as for the initial and final state interaction contributions to the SSA.}}
\begin{ruledtabular}
\begin{tabular}{|l|c|c|c|c|c|c|}
& $p_1^\mu p_1^\nu$ & $k_1^\mu k_1^\nu$ & $k_2^\mu k_2^\nu$ & $p_1^\mu k_1^\nu$ & $p_1^\mu k_2^\nu$ & $k_1^\mu k_2^\nu$  \\
\hline 
$C_u$ & $~~C_F~~$
&$~~C_F~~$ &$~~C_F~~$ & $~-\frac{1}{2N_c}~~$ & $\frac{1}{4}(2C_A-C_F)$
&$~-\frac{1}{4}C_F~~$\\
\hline 
$C_I$  & $C_F$ & $C_F$
       &$C_F$ & $-\frac{1}{2N_c}$&
$\frac{1}{2}$& $-1$\\
\hline 
$C_{F1}$ & $C_F$ & $C_F$
       &$C_F$ & $-\frac{1}{2N_c}$  &
$\frac{C_A}{2}-\frac{1}{N_c^2-2}$&$-\frac{1}{N_c^2-2}$\\
\hline 
$C_{F2}$ & $C_F$ & $C_F$
       &$C_F$ & $-\frac{1}{2N_c}$& $ -\frac{1}{N_c}$&
$C_F-\frac{1}{2N_c}$
 \end{tabular}
\end{ruledtabular}
\label{qqcolor}
\end{table}

We can extend the above calculations to the final state interaction contributions associated with $k_1$ and $k_2$. We summarize the relevant color factors for the different terms in Table~\ref{qqcolor}. The total contribution can be written as
\begin{eqnarray}
{\cal H}_{\textrm{twist-3}}^{\beta (C_{F1}+C_{F2})}
({qq'\to qq'})&=&H_{qq'\to qq'}^{\rm Sivers}\frac{\alpha_s}{2\pi^2}\frac{-q_\perp^\beta}{(q_\perp^2)^2}\left[
2C_F\ln\frac{Q^2}{q_\perp^2}+C_F\left(\ln\frac{1}{R_1^2}+\ln\frac{1}{R_2^2}\right)\right]
 \ , \label{qqsoft1}
\end{eqnarray}
where $H_{qq'\to qq'}^{\rm Sivers}$ was defined in Eq.~(\ref{hqq'}). Clearly, the first term contributes to the leading double logarithms of the SSA for this process. In addition, the divergence associated with the two final state quark jets has the desired structure. Combining the above result with the collinear gluon radiation contributions from incoming partons, see Eq.~(\ref{e30}), we obtain the spin-dependent differential cross section for the ${qq'\to qq'}$ channel 
\begin{eqnarray}
\frac{d\Delta\sigma(S_\perp)}{d\Omega d^2 q_\perp}&=&-H_{qq'\to qq'}^{\rm Sivers}\epsilon^{\alpha\beta}S_\perp^\alpha \frac{\alpha_s}{2\pi^2}\frac{q_\perp^\beta}{(q_\perp^2)^2}x_1x_2\nonumber\\
&&\times \left\{
f_q(x_1){\cal P}_{q'g\to q'g}^{T(<)}\otimes T_{Fq'}(x_2,x_2)
+T_{Fq'}(x_2,x_2){\cal P}_{q\to q}^{(<)}\otimes f_{q}(x_1)\right.\nonumber\\
&&\left.+f_q(x_1)T_{Fq'}(x_2,x_2)\left[
2C_F\ln\frac{Q^2}{q_\perp^2}+C_F\left(\ln\frac{1}{R_1^2}+\ln\frac{1}{R_2^2}\right)\right]\right\} \ ,
\end{eqnarray}
in the correlation limit of $q_\perp\ll P_T$. When taking the Fourier transform to $b_\perp$-space and adding the virtual and jet contributions to cancel the divergences, we expect to obtain the following one-loop result for $W^{T\beta}$ at $\rm LLA'$ accuracy:
\begin{eqnarray}
W_{qq'\to qq'}^{T\beta (1)}&=&H_{qq'\to qq'}^{\rm Sivers}\frac{ib_\perp^\beta}{2}\frac{\alpha_s}{2\pi}x_1x_2\left\{-\ln\frac{\mu^2b_\perp^2}{b_0^2}\Big[
f_q(x_1,\mu){\cal P}_{q'g\to q'g}^T\otimes T_{Fq'}(x_2,x_2,\mu)\right.\nonumber \\
&+&T_{Fq'}(x_2,x_2,\mu){\cal P}_{a'\to q}\otimes f_{a'}(x_1,\mu)\Big]\nonumber\\
&+&\left.
f_q(x_1,\mu)T_{Fq'}(x_2,x_2,\mu)C_F\left[\ln^2\left(\frac{Q^2b_\perp^2}{b_0^2}\right)-\left(\frac{3}{2}-\ln\frac{1}{R_1^2}-\ln\frac{1}{R_2^2}\right)\ln\frac{Q^2b_\perp^2}{b_0^2}\right]\right\} \ . \nonumber \\  
\end{eqnarray}
Here we have included the subtraction of the collinear divergences, and ${\cal P}_{qg\to qg}^T$ and 
${\cal P}_{a'\to q}$ are the complete splitting kernels for the associated twist-three and leading-twist parton distributions, respectively. To derive the above results, we have assumed that the virtual contributions cancel the soft divergences of the real gluon radiation. It is important to show that this is indeed the case. However, this is beyond the scope of this work  and we plan to come back to this issue in a later publication. 

We summarize three important aspects of the above result at this order. First, the collinear splitting is associated with the twist-three and twist-two parton distribution functions. This is the essence of the factorization of collinear gluon radiations as demonstrated in Ref.~\cite{Qiu:2007ey}. Second, the result for the leading double logarithms is consistent with the collinear and soft factorization at $\rm LLA'$. Each of the incoming quark lines contributes half of this double logarithmic term. This is an important feature for the Sudakov resummation. Third, the logarithms associated with the jets are also factorized in terms of the individual jets and we obtain the expected color charges of the final state jets. These results are consistent with the factorization argument. 

In the following two subsections, we will extend the above analysis to two other important channels, $gq\to gq$ and $\bar q q\to gg$. The former is the dominant channel for the SSA in  dijet production for the typical kinematics at RHIC.

\subsubsection{$qg\to qg$}

The leading order derivation for the channel $qg\to qg$ was carried out in Ref.~\cite{Qiu:2007ey}. In order to simplify the analysis for the soft gluon radiation contribution, we follow Ref.~\cite{Sun:2015doa} and decompose the fundamental partonic scattering amplitude as
\begin{equation}
A_1\bar u_j T_{jk}^aT_{ki}^bu_i+A_2\bar u_jT_{jk}^bT_{ki}^au_i \ ,
\end{equation}
with two different color structures at the Born level. Here, $a$ and $b$ are the color indices for the incoming and outgoing gluons, and $i$ and $j$ for the incoming and outgoing quarks. The amplitudes $A_1$ and $A_2$ depend on the momenta of the two incoming particles, $p_1$ and $p_2$, and on the momenta of the two outgoing particles $k_1$ and $k_2$ for the quarks and gluons, respectively. The single spin asymmetry can be formulated starting from the decomposed amplitude above. For example, the initial state interaction contribution can be derived from the following expression:
\begin{eqnarray}
-if_{cad}\left(A_1\bar u_j T_{jk}^dT_{ki}^bu_i+A_2\bar u_j T_{jk}^bT_{ki}^du_i\right)\left(A_1^*\bar u_{i'} T_{i'k'}^bT_{k'j'}^au_{j'}+A_2\bar u_{i'} T_{i'k'}^a
T_{k'j'}^bu_{j'}\right) \ .
\end{eqnarray}
The color indices $i$ and $i'$ are coupled to the adjoint representation of the twist-three Qiu-Sterman matrix element. Therefore, we can rewrite $u_i\bar u_{i'}$ as $T^c_{ii'}$, and the above result leads to
\begin{eqnarray}
C_I:~{\cal H}_0^I&=&\frac{1}{N_{\textrm{color}}}\left(-if_{cad}\right){\rm Tr}\left[\left(A_1T^dT^b+A_2T^bT^d\right)T^c\left(A_1^* T^bT^a+A_2^*T^aT^b\right)\right]\nonumber\\
&=&\frac{1}{N_{\textrm{color}}}\left[-(A_1+A_2)^2+N_c^2A_2^2\right] \ ,
\end{eqnarray}
where $N_{\textrm{color}}=(N_c^2-1)N_cC_F$. Similarly, for the final state interaction with the gluon line, we have
\begin{eqnarray}
C_{F1}:~{\cal H}_0^{F1}&=&\frac{1}{N_{\textrm{color}}}\left(-if_{cbd}\right){\rm Tr}\left[\left(A_1T^aT^d+A_2T^dT^a\right)T^c\left(A_1^* T^bT^a+A_2^*T^aT^b\right)\right]\nonumber\\
&=&\frac{1}{N_{\textrm{color}}}\left[-(A_1+A_2)^2+N_c^2A_1^2\right] \ .
\end{eqnarray}
For the final state interaction contribution from the final state quark line, we have
\begin{eqnarray}
C_{F2}:~{\cal H}_0^{F2}&=&\frac{1}{N_{\textrm{color}}}{\rm Tr}\left[\left(A_1T^cT^aT^b+A_2T^cT^bT^a\right)T^c\left(A_1^* T^bT^a+A_2^*T^aT^b\right)\right]\nonumber\\
&=&\frac{1}{N_{\textrm{color}}}\left[\frac{1}{N_c^2}(A_1+A_2)^2+2A_1A_2^* \right]\ .
\end{eqnarray}
Noticing that the initial state interaction carries a minus sign, the total contribution will be 
\begin{eqnarray}
H_{qg\to qg}^{\rm Sivers}&=&{\cal H}_0^{F1}+{\cal H}_0^{F2}-{\cal H}_0^{I}\nonumber\\
&=&\frac{1}{N_{\textrm{color}}}\left[N_c^2\left(A_1^2-A_2^2\right)+\frac{1}{N_c^2}(A_1+A_2)^2+2A_1A_2^*\right] \ .\label{e72}
\end{eqnarray}
From our parameterization of the Born amplitude, we have
\begin{eqnarray}
&&(A_1+A_2)^2=\frac{2(\hat s^2+\hat u^2)}{-\hat s\hat u},~~A_1A_2^*=-\frac{2(\hat s^2+\hat u^2)}{\hat t^2}\ , \\
&&A_1^2=\frac{2\hat s}{-\hat u}\frac{\hat s^2+\hat u^2}{\hat t^2},~~
A_2^2=\frac{2\hat u}{-\hat s}\frac{\hat s^2+\hat u^2}{\hat t^2}.
\end{eqnarray}
Substituting the above expressions into Eq.~(\ref{e72}), we reproduce the result for the leading order $H_{qg\to qg}^{\rm Sivers}$ in Ref.~\cite{Qiu:2007ey}.

Now, let us turn to the soft gluon radiation contributions. For the initial state interactions, we have soft gluon radiation contributions from the incoming gluon line, the outgoing gluon line and quark line. The relevant diagrams will follow those in Fig.~\ref{gluonradiation0}. In this case, the upper two lines are gluons. The associated amplitudes for these three diagrams are given by
\begin{eqnarray}
&&\frac{2p_1^\mu}{2p_1\cdot k_g}(-if_{cae})(-if_{def})\left(A_1\bar uT^fT^bu+A_2\bar uT^bT^f u\right) \ ,\\
&&\frac{2k_1^\mu}{2k_1\cdot k_g}(-if_{dae})(-if_{cbf})\left(A_1\bar uT^eT^fu+A_2\bar uT^fT^e u\right) \ ,\\
&&\frac{2k_2^\mu}{2k_2\cdot k_g}(-if_{dae})\left(A_1\bar uT^cT^eT^bu+A_2\bar uT^cT^bT^e u\right) \ .
\end{eqnarray}
Here $c$ is the color index of the radiated gluon and $d$ corresponds to the longitudinal gluon from the polarized nucleon attached to the partonic scattering part. The contributions to the SSA come from the interference of the above amplitudes and those in Fig.~\ref{unpolarized}, which we list here for the $qg\to qg$ channel,
\begin{eqnarray}
&&\frac{2p_1^\nu}{2p_1\cdot k_g}(-if_{cag})\left(A_1\bar uT^gT^bu+A_2\bar uT^bT^g u\right) \ ,\\
&&\frac{2k_1^\nu}{2k_1\cdot k_g}(-if_{cbg})\left(A_1\bar uT^aT^gu+A_2\bar uT^gT^a u\right) \ ,\\
&&\frac{2k_2^\nu}{2k_2\cdot k_g}\left(A_1\bar uT^cT^aT^bu+A_2\bar uT^cT^bT^a u\right) \ .
\end{eqnarray}
Similar to the previous case, the following interference terms are simple,
\begin{eqnarray}
p_1^\mu p_1^\nu \Rightarrow C_A, ~~k_1^\mu k_1^\nu\Rightarrow C_A,~~k_2^\mu k_2^\nu\Rightarrow C_F \ , \label{simpleone}
\end{eqnarray}
which will be multiplied by the leading order initial state interaction contribution of Eqs.~(\ref{p1p1})-(\ref{k2k2}). The other interference terms are a little bit more complicated. We have
\begin{eqnarray}
p_1^\mu k_1^\nu &\Rightarrow& \frac{1}{N_{\textrm{color}}}(-if_{cae})(-if_{def})(-if_{cbg}){\rm Tr}\left[\left(A_1T^fT^b+A_2T^bT^f\right)T^d\left(A_1^* T^gT^a+A_2^*T^aT^g\right)\right]\nonumber\\
&&=\frac{N_c}{N_{\textrm{color}}}\left(A_1^2+A_1A_2^*-\frac{N_c^2}{2}A_2^2\right) \ ,\\
p_1^\mu k_2^\nu &\Rightarrow& \frac{1}{N_{\textrm{color}}}(-if_{cae})(-if_{def}){\rm Tr}\left[\left(A_1T^fT^b+A_2T^bT^f\right)T^d\left(A_1^* T^bT^aT^c+A_2^*T^aT^bT^c\right)\right]\nonumber\\
&&=\frac{N_c}{2N_{\textrm{color}}}\left(A_1^2+A_2^2\right) \ ,\\
k_1^\mu k_2^\nu &\Rightarrow& \frac{1}{N_{\textrm{color}}}(-if_{dae})(-if_{cbf}){\rm Tr}\left[\left(A_1T^eT^f+A_2T^fT^e\right)T^d\left(A_1^* T^bT^aT^c+A_2^*T^aT^bT^c\right)\right]\nonumber\\
&&=\frac{N_c}{2N_{\textrm{color}}}\left(-A_1^2-(N_c^2-1)A_2^2+2A_1A_2^*\right) \ .
\end{eqnarray}
By combining them with the associated kinematic contributions in Eqs.~(\ref{p1p1})-(\ref{k1k2}) and adding them, we obtain the leading logarithmic result for the SSA from the initial state interaction for this channel. In particular, the above three terms add up to
\begin{eqnarray}
-C_A{\cal H}_0^I=\frac{C_A}{N_{\textrm{color}}}\left[(A_1+A_2)^2-N_c^2A_2^2\right] \ ,
\end{eqnarray}
which exactly cancels the leading logarithmic contribution from the $p_1^\mu p_1^\nu$ term which also has the color factor $C_A$. We thus obtain the final result for the initial state interaction contribution:
\begin{eqnarray}
{\cal H}_{\textrm{twist-3}}^{\beta(C_I)}&=&-{\cal H}_0^I\frac{\alpha_s}{2\pi^2}\frac{-q_\perp^\beta}{(q_\perp^2)^2}\left[(C_A+C_F)\ln\frac{Q^2}{q_\perp^2}+C_A\ln\frac{1}{R_1^2}+C_F\ln\frac{1}{R_2^2}\right] \ .
\end{eqnarray}
For the final state interaction associated with the final state gluon jet, we have the following amplitudes for the associated diagrams in Fig.~\ref{gluonradiation1}:
\begin{eqnarray}
&&\frac{2p_1^\mu}{2p_1\cdot k_g}(-if_{cae})(-if_{dbf})\left(A_1\bar uT^eT^fu+A_2\bar uT^fT^e u\right) \ ,\\
&&\frac{2k_1^\mu}{2k_1\cdot k_g}(-if_{cbe})(-if_{def})\left(A_1\bar uT^aT^fu+A_2\bar uT^fT^a u\right) \ ,\\
&&\frac{2k_2^\mu}{2k_2\cdot k_g}(-if_{dbe})\left(A_1\bar uT^cT^aT^eu+A_2\bar uT^cT^eT^a u\right) \ .
\end{eqnarray}
Again, the interference contributions from $p_1^\mu p_1^\nu$, $k_1^\mu k_1^\nu$ and $k_2^\mu k_2^\nu$ have the same structure as those for the initial state interaction diagrams in Eq.~(\ref{simpleone}). The remaining contributions can be written as follows:
\begin{eqnarray}
p_1^\mu k_1^\nu &\Rightarrow& \frac{1}{N_{\textrm{color}}}(-if_{cae})(-if_{dbf})(-if_{cbg}){\rm Tr}\left[\left(A_1T^eT^f+A_2T^fT^e\right)T^d\left(A_1^* T^gT^a+A_2^*T^aT^g\right)\right]\nonumber\\
&&=\frac{N_c}{N_{\textrm{color}}}\left(A_2^2+A_1A_2^*-\frac{N_c^2}{2}A_1^2\right) \ ,\\
p_1^\mu k_2^\nu &\Rightarrow& \frac{1}{N_{\textrm{color}}}(-if_{cae})(-if_{dbf}){\rm Tr}\left[\left(A_1T^eT^f+A_2T^fT^e\right)T^d\left(A_1^* T^bT^aT^c+A_2^*T^aT^bT^c\right)\right]\nonumber\\
&&=\frac{N_c}{2N_{\textrm{color}}}\left(2A_1A_2^*-A_2^2-(N_c^2-1)A_1^2\right) \ ,\\
k_1^\mu k_2^\nu &\Rightarrow& \frac{1}{N_{\textrm{color}}}(-if_{cbe})(-if_{def}){\rm Tr}\left[\left(A_1T^aT^f+A_2T^fT^a\right)T^d\left(A_1^* T^bT^aT^c+A_2^*T^aT^bT^c\right)\right]\nonumber\\
&&=\frac{N_c}{2N_{\textrm{color}}}\left(A_1^2+A_2^2\right) \ .
\end{eqnarray}
Adding up the three contributions, we have 
\begin{eqnarray}
-C_A{\cal H}_0^{F1}=\frac{C_A}{N_{\textrm{color}}}\left[(A_1+A_2)^2-N_c^2A_1^2\right] \ .
\end{eqnarray}
The total contribution from the final state interaction with the gluon jet is given by
\begin{eqnarray}
{\cal H}_{\textrm{twist-3}}^{\beta(C_{F1})}&=&{\cal H}_0^{F1}\frac{\alpha_s}{2\pi^2}\frac{-q_\perp^\beta}{(q_\perp^2)^2}\left[(C_A+C_F)\ln\frac{Q^2}{q_\perp^2}+C_A\ln\frac{1}{R_1^2}+C_F\ln\frac{1}{R_2^2}\right] \ .
\end{eqnarray}
For the final state interaction with the quark line, we find the following amplitudes for the associated diagrams of Fig.~\ref{gluonradiation2}:
\begin{eqnarray}
&&\frac{2p_1^\mu}{2p_1\cdot k_g}(-if_{cae})\left(A_1\bar uT^dT^eT^bu+A_2\bar uT^dT^bT^e u\right) \ ,\\
&&\frac{2k_1^\mu}{2k_1\cdot k_g}(-if_{cbe})\left(A_1\bar uT^dT^aT^eu+A_2\bar uT^dT^eT^a u\right) \ ,\\
&&\frac{2k_2^\mu}{2k_2\cdot k_g}\left(A_1\bar uT^cT^dT^aT^bu+A_2\bar uT^cT^dT^bT^a u\right) \ ,
\end{eqnarray}
and the interference terms are
\begin{eqnarray}
p_1^\mu k_1^\nu &\Rightarrow& \frac{1}{N_{\textrm{color}}}(-if_{cae})(-if_{cbg}){\rm Tr}\left[\left(A_1T^dT^eT^b+A_2T^dT^bT^e\right)T^d\left(A_1^* T^gT^a+A_2^*T^aT^g\right)\right]\nonumber\\
&&=\frac{N_c}{2N_{\textrm{color}}}\left(-A_1^2-A_2^2-4A_1A_2^*\right) \ ,\\
p_1^\mu k_2^\nu &\Rightarrow& \frac{1}{N_{\textrm{color}}}(-if_{cae}){\rm Tr}\left[\left(A_1T^dT^eT^b+A_2T^dT^bT^e\right)T^d\left(A_1^* T^bT^aT^c+A_2^*T^aT^bT^c\right)\right]\nonumber\\
&&=\frac{1}{2N_cN_{\textrm{color}}}\left(-2A_1A_2^*-A_1^2+(N_c^2-1)A_2^2\right) \ ,\\
k_1^\mu k_2^\nu &\Rightarrow& \frac{1}{N_{\textrm{color}}}(-if_{cbe}){\rm Tr}\left[\left(A_1T^dT^aT^e+A_2T^dT^eT^a\right)T^d\left(A_1^* T^bT^aT^c+A_2^*T^aT^bT^c\right)\right]\nonumber\\
&&=\frac{1}{2N_cN_{\textrm{color}}}\left(-2A_1A_2^*-A_2^2+(N_c^2-1)A_1^2\right) \ .
\end{eqnarray}
The above three terms lead to 
\begin{eqnarray}
-C_A{\cal H}_0^{F2}=\frac{C_A}{N_{\textrm{color}}}\left[-\frac{1}{N_c^2}(A_1+A_2)^2-2A_1A_2^*\right] \ .
\end{eqnarray}
The final result from the final state interaction with the quark line is
\begin{eqnarray}
{\cal H}_{\textrm{twist-3}}^{\beta(C_{F2})}&=&{\cal H}_0^{F2}\frac{\alpha_s}{2\pi^2}\frac{-q_\perp^\beta}{(q_\perp^2)^2}\left[(C_A+C_F)\ln\frac{Q^2}{q_\perp^2}+C_A\ln\frac{1}{R_1^2}+C_F\ln\frac{1}{R_2^2}\right] \ .
\end{eqnarray}
Adding up all initial and final state interaction contributions, we obtain the following result for the SSA at the leading logarithmic order:
\begin{eqnarray}
{\cal H}_{\textrm{twist-3}}^{\beta}(gq\to gq)&=&H_{qg\to qg}^{\rm Sivers}\frac{\alpha_s}{2\pi^2}\frac{-q_\perp^\beta}{(q_\perp^2)^2}\left[(C_A+C_F)\ln\frac{Q^2}{q_\perp^2}+C_A\ln\frac{1}{R_1^2}+C_F\ln\frac{1}{R_2^2}\right] \ .
\end{eqnarray}
Including the collinear gluon radiation contributions from the incoming partons as in Eq.~(\ref{e30}), we obtain the spin-dependent differential cross section for the $gq\to gq$ channel in the correlation limit of $q_\perp\ll P_T$ which is given by
\begin{eqnarray}
\frac{d\Delta\sigma(S_\perp)}{d\Omega d^2 q_\perp}&=&-H_{qg\to qg}^{\rm Sivers}\epsilon^{\alpha\beta}S_\perp^\alpha \frac{\alpha_s}{2\pi^2}\frac{q_\perp^\beta}{(q_\perp^2)^2}x_1x_2\nonumber\\
&&\times \left\{
f_g(x_1){\cal P}_{qg\to qg}^{T(<)}\otimes T_{Fq}(x_2,x_2)
+T_{Fq}(x_2,x_2){\cal P}_{g\to g}^{(<)}\otimes f_{g}(x_1)\right.\nonumber\\
&&\left.+f_g(x_1)T_{Fq}(x_2,x_2)\left[
(C_A+C_F)\ln\frac{Q^2}{q_\perp^2}+C_A\ln\frac{1}{R_1^2}+C_F\ln\frac{1}{R_2^2}\right]\right\} \ .
\end{eqnarray}
After taking the Fourier transform to $b_\perp$-space, we expect to have the following one-loop result for $W^{T\beta}$:
\begin{eqnarray}
W_{gq\to gq}^{T\beta (1)}&=&H_{qg\to qg}^{\rm Sivers}\frac{ib_\perp^\beta}{2}\frac{\alpha_s}{2\pi}x_1x_2\left\{-\ln\frac{\mu^2b_\perp^2}{b_0^2}\Big[f_g(x_1,\mu)
{\cal P}_{qg\to qg}^T\otimes T_{Fq}(x_2,x_2,\mu)\right. \nonumber \\
&+&T_{Fq}(x_2,x_2,\mu){\cal P}_{a'\to g}\otimes f_{a'}(x_1,\mu)\Big]\nonumber\\
&+&
f_g(x_1,\mu)T_{Fq}(x_2,x_2,\mu)\left[\frac{C_A+C_F}{2}\ln^2\left(\frac{Q^2b_\perp^2}{b_0^2}\right)\right.\nonumber\\
&&~~~\left.\left.-\left(\frac{3}{2}C_F+2\beta_0C_A-C_A\ln\frac{1}{R_1^2}-C_F\ln\frac{1}{R_2^2}\right)\ln\frac{Q^2b_\perp^2}{b_0^2}\right]\right\} \ .
\end{eqnarray}

\subsubsection{$q\bar q\to gg$}

The computation for the $q\bar q\to gg$ channel is very similar to the case discussed above. Here, we can parametrize the leading Born amplitude as
\begin{eqnarray}
A_1\bar vT^aT^b u +A_2\bar v T^bT^a u \ ,
\end{eqnarray}
where, of course, the amplitudes will be different from those for the $qg\to qg$ channel. In particular, we have $(A_1+A_2)^2=\frac{2(\hat t^2+\hat u^2)}{\hat t\hat u}$, $A_1A_2^*=\frac{2(\hat t^2+\hat u^2)}{\hat s^2}$, $A_1^2=\frac{2\hat t}{\hat u}\frac{2(\hat t^2+\hat u^2)}{\hat s^2}$ and $A_2^2=\frac{2\hat u}{\hat t}\frac{2(\hat t^2+\hat u^2)}{\hat s^2}$ for the current case.

The single spin asymmetry comes from the initial state interaction with the antiquark line and the two final state gluon lines. For the initial state interaction contribution, we have
\begin{eqnarray}
C_I^{(q\bar q)}:~{\cal H}_{0q\bar q}^I&=&\frac{1}{N_{\textrm{color}}^{(q\bar q)}}{\rm Tr}\left[\left(A_1T^cT^aT^b+A_2T^cT^bT^a\right)T^c\left(A_1^* T^bT^a+A_2^*T^aT^b\right)\right]\nonumber\\
&=&\frac{1}{N_{\textrm{color}}^{(q\bar q)}}\left[\frac{1}{N_c^2}(A_1+A_2)^2+2A_1A_2^*\right] \ ,
\end{eqnarray}
where $N_{\textrm{color}}^{(q\bar q)}=N_c^2C_F$. For the final state interaction contribution with the gluon line of $k_1$, we obtain
\begin{eqnarray}
C_{F1}^{(q\bar q)}:~{\cal H}_{0q\bar q}^{F1}&=&\frac{1}{N_{\textrm{color}}^{(q\bar q)}}\left(-if_{cae}\right){\rm Tr}\left[\left(A_1T^eT^b+A_2T^bT^e\right)T^c\left(A_1^* T^bT^a+A_2^*T^aT^b\right)\right]\nonumber\\
&=&\frac{1}{N_{\textrm{color}}^{(q\bar q)}}\left[{N_c^2}A_2^2-(A_1+A_2)^2\right] \ .
\end{eqnarray}
Using the the symmetry of the channel, the final state interaction contribution from the gluon line $k_2$ can be obtained from the above result as
\begin{eqnarray}
C_{F2}^{(q\bar q)}:~{\cal H}_{0q\bar q}^{F2}
&=&\frac{1}{N_{\textrm{color}}^{(q\bar q)}}\left[{N_c^2}A_1^2-(A_1+A_2)^2\right] \ .
\end{eqnarray}
The total contribution at the leading order is given by
\begin{eqnarray}
H_{q\bar q\to gg}^{\rm Sivers}&=&{\cal H}_{0q\bar q}^{F1}+{\cal H}_{0q\bar q}^{F2}-{\cal H}_{0q\bar q}^{I}\nonumber\\
&=&\frac{1}{N_{\textrm{color}}^{(q\bar q)}}\left[N_c^2\left(A_1^2+A_2^2\right)-2A_1A_2^*-\frac{2N_c^2+1}{N_c^2}(A_1+A_2)^2\right] \ .\label{loqqbar}
\end{eqnarray}
Substituting the expressions of $A_1$ and $A_2$ into the above equation, we reproduce the result for the leading order $H_{q\bar q\to gg}^{\rm Sivers}$ in Ref.~\cite{Qiu:2007ey}.

Next, we can work out the soft gluon radiation contribution as well. For the initial state interaction contributions ($C_I$), we consider the soft gluon radiation from the incoming anti-quark line, and two the outgoing gluon lines. The relevant diagrams again correspond to those shown in Fig.~\ref{gluonradiation0}. The associated amplitudes for these three diagrams are
\begin{eqnarray}
&&\frac{2p_1^\mu}{2p_1\cdot k_g}\left(A_1\bar vT^cT^dT^aT^bu+A_2\bar vT^cT^dT^bT^a u\right) \ ,\\
&&\frac{2k_1^\mu}{2k_1\cdot k_g}(-if_{cae})\left(A_1\bar vT^dT^eT^bu+A_2\bar vT^dT^bT^e u\right) \ ,\\
&&\frac{2k_2^\mu}{2k_2\cdot k_g}(-if_{cbe})\left(A_1\bar vT^dT^aT^eu+A_2\bar vT^dT^eT^a u\right) \ ,
\end{eqnarray}
where $c$ is again the color index of the radiated gluon and $d$ corresponds to the longitudinal gluon from the polarized nucleon. The single spin asymmetry contributions come from the interference of the above amplitudes and those of Fig.~\ref{unpolarized}. For $q\bar q\to gg$ channel, we have
\begin{eqnarray}
&&\frac{2p_1^\nu}{2p_1\cdot k_g}\left(A_1\bar vT^cT^aT^bu+A_2\bar vT^cT^bT^a u\right) \ ,\\
&&\frac{2k_1^\nu}{2k_1\cdot k_g}(-if_{cag})\left(A_1\bar vT^gT^bu+A_2\bar vT^bT^g u\right) \ ,\\
&&\frac{2k_2^\nu}{2k_2\cdot k_g}(-if_{cbg})\left(A_1\bar vT^aT^gu+A_2\bar vT^gT^a u\right) \ .
\end{eqnarray}
In our previous notation, the resulting interference terms are
\begin{eqnarray}
p_1^\mu p_1^\nu \Rightarrow C_F, ~~k_1^\mu k_1^\nu\Rightarrow C_A,~~k_2^\mu k_2^\nu\Rightarrow C_A \ .
\end{eqnarray}
The other interference terms can be evaluated as
\begin{eqnarray}
p_1^\mu k_1^\nu &\Rightarrow& \frac{1}{N_{\textrm{color}}^{(q\bar q)}}(if_{cae}){\rm Tr}\left[\left(A_1T^cT^dT^aT^b+A_2T^cT^dT^bT^a\right)T^d\left(A_1^* T^bT^e+A_2^*T^eT^b\right)\right]\nonumber\\
&&=\frac{1}{2N_cN_{\textrm{color}}^{(q\bar q)}}\left((N_c^2-1)A_2^2-A_1^2-2A_1A_2^*\right) \ ,\\
p_1^\mu k_2^\nu &\Rightarrow& \frac{1}{N_{\textrm{color}}^{(q\bar q)}}(if_{cbe}){\rm Tr}\left[\left(A_1T^cT^dT^aT^b+A_2T^cT^dT^bT^a\right)T^d\left(A_1^* T^eT^a+A_2^*T^aT^e\right)\right]\nonumber\\
&&=\frac{1}{2N_cN_{\textrm{color}}^{(q\bar q)}}\left((N_c^2-1)A_1^2-A_2^2-2A_1A_2^*\right) \ ,\\
k_1^\mu k_2^\nu &\Rightarrow& \frac{1}{N_{\textrm{color}}^{(q\bar q)}}(-if_{cae})(if_{cbf}){\rm Tr}\left[\left(A_1T^dT^eT^b+A_2T^dT^bT^e\right)T^d\left(A_1^* T^fT^a+A_2^*T^aT^f\right)\right]\nonumber\\
&&=\frac{N_c}{N_{\textrm{color}}^{(q\bar q)}}\left(-\frac{1}{2}(A_1^2+A_2^2)-2A_1A_2^*\right) \ .
\end{eqnarray}
The above three terms add up to
\begin{eqnarray}
-C_A{\cal H}_{0q\bar q}^I=\frac{1}{N_{\textrm{color}}^{(q\bar q)}}\left[-\frac{1}{N_c}(A_1+A_2)^2-2N_cA_1A_2^*\right] \ ,
\end{eqnarray}
which exactly cancels the leading log contribution from $k_1^\mu k_1^\nu$ and $k_2^\mu k_2^\nu$ with color factor $C_A$. Therefore, we obtain the following final result for the initial state interaction contribution:
\begin{eqnarray}
{\cal H}_{\textrm{twist-3}}^{\beta(C_I)}=-{\cal H}_{0q\bar q}^I\frac{\alpha_s}{2\pi^2}\frac{-q_\perp^\beta}{(q_\perp^2)^2}\left[C_F2\ln\frac{Q^2}{q_\perp^2}+C_A\left(\ln\frac{1}{R_1^2}+\ln\frac{1}{R_2^2}\right)\right] \ .
\end{eqnarray}
For the final state interaction associated with the final state gluon jet $k_1$, we find the following amplitudes, see Fig.~\ref{gluonradiation1},
\begin{eqnarray}
&&\frac{2p_1^\mu}{2p_1\cdot k_g}(-if_{dae})\left(A_1\bar vT^cT^eT^bu+A_2\bar vT^cT^bT^e u\right) \ ,\\
&&\frac{2k_1^\mu}{2k_1\cdot k_g}(-if_{caf})(-if_{dfe})\left(A_1\bar vT^eT^bu+A_2\bar vT^bT^e u\right) \ ,\\
&&\frac{2k_2^\mu}{2k_2\cdot k_g}(-if_{cbf})(-if_{dae})\left(A_1\bar vT^eT^fu+A_2\bar vT^fT^e u\right) \ .
\end{eqnarray}
Again, the interference contributions from $p_1^\mu p_1^\nu$, $k_1^\mu k_1^\nu$ and $k_2^\mu k_2^\nu$ have the same structure as those for the initial state interaction diagrams. The remaining contributions can be written as
\begin{eqnarray}
p_1^\mu k_1^\nu &\Rightarrow& \frac{1}{N_{\textrm{color}}^{(q\bar q)}}(-if_{dae})(if_{cag}){\rm Tr}\left[\left(A_1T^cT^eT^b+A_2T^cT^bT^e\right)T^d\left(A_1^* T^bT^g+A_2^*T^gT^b\right)\right]\nonumber\\
&&=\frac{N_c}{2N_{\textrm{color}}^{(q\bar q)}}\left(A_1^2+A_2^2\right) \ ,\\
p_1^\mu k_2^\nu &\Rightarrow& \frac{1}{N_{\textrm{color}}^{(q\bar q)}}(-if_{dae})(if_{cbg}){\rm Tr}\left[\left(A_1T^cT^eT^b+A_2T^cT^bT^e\right)T^d\left(A_1^* T^gT^a+A_2^*T^aT^g\right)\right]\nonumber\\
&&=\frac{N_c}{2N_{\textrm{color}}^{(q\bar q)}}\left(-A_1^2+2A_1A_2^*-(N_c^2-1)A_2^2\right) \ ,\\
k_1^\mu k_2^\nu &\Rightarrow& \frac{1}{N_{\textrm{color}}^{(q\bar q)}}(-if_{caf})(-if_{dfe})(if_{cbg}){\rm Tr}\left[\left(A_1T^eT^b+A_2T^bT^e\right)T^d\left(A_1^* T^gT^a+A_2^*T^aT^g\right)\right]\nonumber\\
&&=\frac{N_c}{2N_{\textrm{color}}^{(q\bar q)}}\left(2A_1^2+2A_1A_2^*-N_c^2A_2^2\right) \ .
\end{eqnarray}
Adding the three terms above, we find 
\begin{eqnarray}
-C_A{\cal H}_{0q\bar q}^{F1}=\frac{C_A}{N_{\textrm{color}}^{(q\bar q)}}\left[(A_1+A_2)^2-N_c^2A_2^2\right] \ .
\end{eqnarray}
The total contribution from the final state interaction with the gluon jet leads to
\begin{eqnarray}
{\cal H}_{\textrm{twist-3}}^{\beta(C_{F1})}={\cal H}_{0q\bar q}^{F1}\frac{\alpha_s}{2\pi^2}\frac{-q_\perp^\beta}{(q_\perp^2)^2}\left[2C_F\ln\frac{Q^2}{q_\perp^2}+C_A\left(\ln\frac{1}{R_1^2}+\ln\frac{1}{R_2^2}\right)\right] \ .
\end{eqnarray}
Again, from the symmetry under interchange of the final state particles, we can obtain the soft gluon radiation contribution for the final state interaction with the gluon line $k_2$. Adding all these contributions, we have the following result for the SSA at the leading logarithmic order,
\begin{eqnarray}
{\cal H}_{\textrm{twist-3}}^{\beta}(q\bar q\to gg)=H_{q\bar q\to gg}^{\rm Sivers}\frac{\alpha_s}{2\pi^2}\frac{-q_\perp^\beta}{(q_\perp^2)^2}\left[2C_F\ln\frac{Q^2}{q_\perp^2}+C_A\left(\ln\frac{1}{R_1^2}+\ln\frac{1}{R_2^2}\right)\right] \ .
\end{eqnarray}
Adding the collinear gluon radiation contribution, we find the following result:
\begin{eqnarray}
\frac{d\Delta\sigma(S_\perp)}{d\Omega d^2 q_\perp}&=&-H_{q\bar q\to gg}^{\rm Sivers}\epsilon^{\alpha\beta}S_\perp^\alpha \frac{\alpha_s}{2\pi^2}\frac{q_\perp^\beta}{(q_\perp^2)^2}x_1x_2\nonumber\\
&&\times \left\{
f_{\bar q}(x_1){\cal P}_{qg\to qg}^{T(<)}\otimes T_{Fq}(x_2,x_2)
+T_{Fq}(x_2,x_2){\cal P}_{a\to \bar q}^{(<)}\otimes f_{\bar q}(x_1)\right.\nonumber\\
&&\left.+f_{\bar q}(x_1)T_{Fq}(x_2,x_2)\left[
2C_F\ln\frac{Q^2}{q_\perp^2}+C_A\left(\ln\frac{1}{R_1^2}+\ln\frac{1}{R_2^2}\right)\right]\right\} \ ,
\end{eqnarray}
for the $q\bar q\to gg$ channel. Taking again the Fourier transform to $b_\perp$-space should lead to the following one-loop result:
\begin{eqnarray}
W_{q\bar q\to gg}^{T\beta (1)}&=&H_{q\bar q\to gg}^{\rm Sivers}\frac{ib_\perp^\beta}{2}\frac{\alpha_s}{2\pi}x_1x_2\left\{-\ln\frac{\mu^2b_\perp^2}{b_0^2}\Big[
f_{\bar q}(x_1,\mu){\cal P}_{qg\to qg}^T\otimes T_{Fq}(x_2,x_2,\mu)\right.\nonumber \\
&+&T_{Fq}(x_2,x_2,\mu){\cal P}_{a\to \bar q}\otimes f_a(x_1,\mu)\Big]\nonumber\\
&+&f_{\bar q}(x_1,\mu)T_{Fq}(x_2,x_2,\mu)\left[C_F\ln^2\left(\frac{Q^2b_\perp^2}{b_0^2}\right)\right.\nonumber\\
&-&\left.\left.\left(\frac{3}{2}C_F-C_A\left(\ln\frac{1}{R_1^2}+\ln\frac{1}{R_2^2}\right)\right)\ln\frac{Q^2b_\perp^2}{b_0^2}\right]\right\}\ .
\end{eqnarray}

\subsection{Factorization and Resummation at $\rm LLA'$}

Our above results for the soft gluon radiation contribution at one-loop order are very instructive for developing 
a factorization argument according to which one can perform the resummation of logarithms in $q_\perp/P_T$.
Three important types of contributions are incorporated and explicitly presented in our calculations: (1) collinear gluon radiation from the incoming hadrons, (2) 
collinear gluon radiation from the final state jets, and (3) the leading double logarithms from soft gluon radiation. 

\begin{figure}[t]
\begin{center}
\includegraphics[width=11cm]{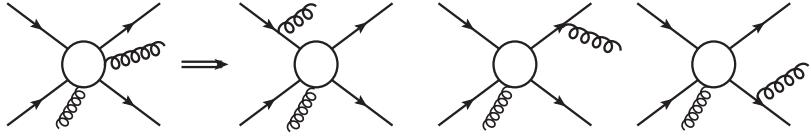}
\end{center}
\vskip -0.4cm \caption{\it Summary of the one-loop calculations of the soft gluon radiation contributions to the SSA in dijet production.}
\label{one-loop-summary}
\end{figure}

We argue that these contributions to the $\rm LLA'$ will continue to factorize at higher orders. As illustrated in Fig.~\ref{one-loop-summary}, we can summarize the above calculations of the one-loop soft gluon radiations in a factorization structure. First, the power counting analysis simplifies the kinematics. As shown in this figure, the initial/final state interactions needed to generate a phase for the SSA in this process can be factorized into a hard partonic scattering amplitude, in analogy to the leading order Born amplitude for the spin-averaged case. Additional soft gluon radiation only appears in association with the external lines. This not only simplifies the detailed derivations at this order, but also shows that the soft gluon radiation associated with the final state jets can be generalized to all orders. 

In our example, we have chosen a physical polarization for the radiated gluon, such that the jet contribution comes only from the squared diagrams where the radiated gluon is attached to one particular jet, e.g. either $k_1$ or $k_2$. These emissions always result in a leading contribution proportional to $1/q_\perp^2\ln(1/R^2)$. An evolution equation can be derived to resum higher order emissions and the final result can be written in terms of the Sudakov resummation form factor of Eq.~(\ref{sudakov-spin}). 

Similar arguments can be made for the collinear gluon radiation associated with the incoming hadrons. These collinear gluon also contribute terms 
of order $1/q_\perp^2$, which will be multiplied by the splitting of the associated parton distribution and/or twist-three Qiu-Sterman matrix element. The all order resummation can be carried out by solving the relevant DGLAP equations. This can be achieved by evaluating the distributions at the scale $\mu_b=b_0/b_\perp$ and by including the associated anomalous dimensions in the Sudakov resummation form factor. 

The leading double logarithms from soft gluon radiation are associated with kinematics overlapping with the collinear gluon radiation from the incoming partons. The resummation of these double logarithms can be carried out by solving the associated Collins-Soper evolution equations for the TMD parton distributions. The double logarithms from two TMD parton distributions give the leading double logarithms for the final differential cross section. This argument has been verified
explicitly in the above derivation. We expect that this can be generalized to higher orders as well and an all order resummation can be obtained in the form of the Sudakov form factor in Eq.~(\ref{sudakov-spin}).

\section{Factorization Breaking Effects at NLL}

To exhibit factorization breaking effects beyond the LLA, we consider the channel $qq'\to qq'$ as an example. From the derivation in the last section, we obtain the soft gluon radiation contribution to the transverse spin dependent differential cross section (see Eqs.~(\ref{p1p1})--(\ref{k1k2}),(\ref{qqsoft}),(\ref{qqsoft1}))
\begin{eqnarray}
{\cal H}_{\textrm{twist-3}}^{\beta}=\frac{\alpha_s}{2\pi^2}\frac{-q_\perp^\beta}{(q_\perp^2)^2}\left\{H_{q_iq_j\to q_iq_j}^{\rm Sivers}\left[
2C_F\ln\frac{Q^2}{q_\perp^2}+C_F\left(\ln\frac{1}{R_1^2}+\ln\frac{1}{R_2^2}\right)\right]
+\widetilde{\Gamma}_{sn}^{(qq')}\right\} \ , \label{qqsoft1a}
\end{eqnarray}
where $h_{q_iq_j\to q_iq_j}^{(0)}=\frac{\alpha_s^2\pi}{\hat s^2}\frac{2(\hat s^2+\hat u^2)}{\hat t^2}$ and 
\begin{equation}
    \widetilde{\Gamma}_{sn}^{(qq')}=h_{q_iq_j\to q_iq_j}^{(0)}\left[-\frac{1}{N_c}\ln\frac{\hat s^2}{\hat t\hat u}-4\ln\frac{\hat t}{\hat u}\right] \ .
\end{equation}
Comparing to Eq.~(66) of Ref.~\cite{Sun:2015doa}, we observe that this term is different from the analogous term in the unpolarized case, implying that it cannot be factorized into a spin independent soft factor for the $qq'\to qq'$ process. 

Even if we consider only {\it initial state} soft gluons for the transverse spin dependent differential cross section, we get a result different from the unpolarized one. Here we have
\begin{eqnarray}
{\cal H}_{\textrm{twist-3}}^{\beta(C_I)}=\frac{\alpha_s}{2\pi^2}\frac{-q_\perp^\beta}{(q_\perp^2)^2}\left\{C_Ih_{q_iq_j\to q_iq_j}^{(0)}\left[
2C_F\ln\frac{Q^2}{q_\perp^2}+C_F\left(\ln\frac{1}{R_1^2}+\ln\frac{1}{R_2^2}\right)\right]
+\widetilde{\Gamma}_{sn}^{(qq',C_I)}\right\} \ ,
\end{eqnarray}
where
\begin{equation}
\widetilde{\Gamma}_{sn}^{(qq',C_I)}=h_{q_iq_j\to q_iq_j}^{(0)}\left[2\ln\frac{\hat s^2}{\hat t\hat u}-C_F\ln\frac{\hat t}{\hat u}\right]\ .
\end{equation}
The same findings occur for the other two channels studied in the previous section, $qg\to qg$ and $q\bar q\to gg$.

\begin{figure}[t]
\begin{center}
\includegraphics[width=5cm]{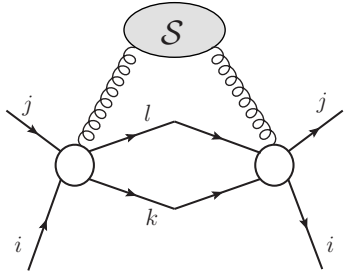}
\end{center}
\vskip -0.4cm \caption{\it Soft gluon radiation for the unpolarized case in dijet production. The soft factor can be constructed from the relevant color space configuration of $ij\to kl$, where $ij$ and $kl$ represent the color indices for incoming and outgoing partons in the partonic $2\to 2$ process. }
\label{unpolarizefactorization}
\end{figure}

Let us recall the factorization argument for soft gluon radiation in dijet production. Following the analysis of the generic soft gluon radiation in this process, see, e.g., in Refs.~\cite{Kidonakis:1998bk,Kidonakis:1998nf,Kidonakis:2000gi}, one can derive the associated soft factor for the TMD factorization in matrix form of the relevant color spaces~\cite{Sun:2014gfa,Sun:2015doa}. As shown in Fig.~\ref{unpolarizefactorization}, the color configuration in the $2\to 2 $ partonic process can be written as $ij\to kl$, where $ij$ and $kl$ represent the color indices for the incoming and outgoing partons, respectively. The independent color space tensors carry these indices. The relevant amplitudes for soft gluon radiation and their complex conjugates are derived based on these configurations as well. In TMD factorization, the associated soft factors are formulated in terms of a matrix form on the basis of these color configurations as well, where the external lines will be replaced by the Wilson lines.  

\begin{figure}[t]
\begin{center}
\includegraphics[width=5cm]{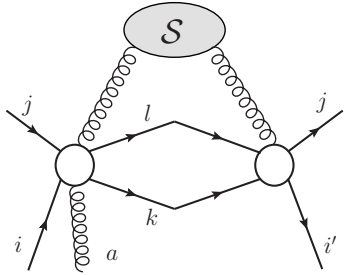}
\end{center}
\vskip -0.4cm \caption{\it Soft gluon radiation for the SSA contributions in the dijet process. The contributions arise as the interference between the amplitude for $ij(a)\to kl$ with an additional gluon attachment from the polarized nucleon and the amplitude without the gluon attachment $i'j\to kl$, where $a$ represents the color index of the attached gluon. The final result is derived by contracting color indices $i$ and $i'$ with $T^a_{ii'}$. Therefore, the color flow is totally different from that in Fig.~\ref{unpolarizefactorization} for the unpolarized case. }
\label{ssafactorization}
\end{figure}

However, for the SSA in the dijet process, an additional gluon attachment from the polarized nucleon is needed to generate the phase required for the Sivers effect. Because of this, the color flow of the partonic process is totally different from that in the unpolarized case. As shown in Fig.~\ref{ssafactorization}, the single spin asymmetry comes from the interference between the amplitude with gluon attachment and that without gluon attachment. The attaching gluon carries color (with adjoint representation index $a$ in this diagram). The color flow for the amplitude shown in this plot can be written as $ij(a)\to kl$ for the left side and $i'j\to kl$ for the right side, respectively. The final result for the SSA is derived by contracting $i$ and $i'$ via the matrix $T^a_{ii'}$ representing the attaching gluon $a$. Because of this additional color flow caused by the gluon attachment, the color space configurations differ from those in the unpolarized case. As a consequence, the soft factor changes, and the result for the spin-averaged case derived in Ref.~\cite{Sun:2014gfa,Sun:2015doa} does not apply here. This is indeed precisely what our calculations demonstrate. 

As just described, the difficulty to apply the standard soft factor factorization for the SSA comes from the fact that we need to generate a phase in order to obtain a nonzero SSA for this process. The phase comes from the imaginary part of the propagator (pole contribution), which may carry a different sign compared to the real part. For unpolarized scattering at this order, in a similar diagram like Fig.~\ref{ssafactorization}, the attached gluon can be incorporated into a relevant gauge link in the  unpolarized quark distribution. This is only possible when there is no pole contribution from the propagators associated with the attached gluons. After this, the color index for the quark line entering the hard part from the left side will be the same as the on right side, i.e., $i\to i'$. As a consequence, we can derive the associated soft factor in TMD factorization as indicated in Fig.~\ref{unpolarizefactorization}. Of course, as mentioned in the Introduction, at higher order ${\cal O}(\alpha_s^3)$ and beyond, the factorization for the unpolarized cross section also breaks down. The reason is the same: one can have two poles in the propagators associated with two gluon attachments, which can not be factorized into the relevant gauge link in the TMD quark distributions; see, examples given in Refs.~\cite{Vogelsang:2007jk,Collins:2007nk}. 

To summarize, the soft gluon radiation contribution to the SSA in the dijet production process in $pp$ collisions cannot be factorized into a spin independent soft factor because of the pole contributions, thus breaking standard TMD factorization. Since the soft factor carries a next-to-leading logarithmic contribution, the factorization for the SSA breaks down at NLL as explicitly shown by the above example of $qq'\to qq'$ channel.
At the leading logarithmic level, we can argue that the relevant soft gluon radiation belongs to the parton distributions and the final state jets. The associated large logarithms can be resummed through the evolution of these parton distributions and jet functions, and factorization is only broken beyond leading logarithmic accuracy. Therefore, the SSA for dijet production in polarized $pp$ collisions may provide a unique opportunity to explore factorization breaking effects. We will study the associated phenomenology in the following section.

We note that it is conceivable that the soft gluon radiation in Fig.~\ref{ssafactorization} may be factorized into a ``non-traditional'' soft factor that is unique to the SSA. This might be achieved by setting up different color tensors on the left side and the right side of the diagram. On the right side of the diagram in Fig.~\ref{ssafactorization} we have a standard $2\to 2$ partonic channel with its simple and standard color flow. On the left side, we have a $3\to 2 $ partonic channel, for which recent developments on the soft gluon evolution in $2\to 3$ processes~\cite{Kyrieleis:2005dt,Sjodahl:2008fz} may become useful. The color-contraction of the two sides is expected to give rise to a non-square matrix problem. It appears likely that the structure of one-loop soft gluon radiation in the single-transverse spin cross section may be understood in this way; whether one can generalize this to higher orders in TMD factorization would be an open question. It is worthwhile to further pursue a study along this direction, which, however, is beyond the scope of the current paper. We hope to address this in a future publication. 

\section{Phenomenological Study and Comparison to STAR Data}

Within the improved leading logarithmic approximation, we can write the differential cross sections for the spin averaged and spin dependent cases in the correlation limit $q_\perp\ll P_T$ as follows:
\begin{eqnarray}
\frac{d^4\sigma}
{d\Omega
d^2q_{\perp}}&=&\sum_{abcd}\int\frac{d^2\vec{b}_\perp}{(2\pi)^2}
e^{i\vec{q}_\perp\cdot
\vec{b}_\perp}W_{ab\to cd}^{uu}(x_1,x_2,b_\perp) \ ,\\
\frac{d\Delta\sigma(S_\perp)}
     {d\Omega d^2\vec{q}_\perp}
&=&\epsilon^{\alpha\beta}S_\perp^\alpha\sum_{abcd}\int\frac{d^2\vec{b}_\perp}{(2\pi)^2}
e^{i\vec{q}_\perp\cdot
\vec{b}_\perp}W_{ab\to cd}^{T\beta}(x_1,x_2,b_\perp)\ .
\end{eqnarray}
Following our arguments on the improved leading logarithmic factorization, the resummation formulas for the above $W^{uu}$ and $W^{T\beta}$ can be written as
\begin{eqnarray}
W_{ab\to cd}^{T\beta}\big|_{\rm LLA'}&=&\frac{ib_\perp^\beta}{2}x_1 f_a(x_1,\mu_b)x_2 T_{Fb}(x_2,\mu_b) H_{ab\to cd}^{\rm Sivers}(P_T,x_1,x_2) e^{-S_{\rm Sud}^T(Q^2,b_\perp)} \ ,\\
W_{ab\to cd}^{uu}\big|_{\rm LLA'}&=&x_1 f_a(x_1,\mu_b)x_2 f_{b}(x_2,\mu_b) H_{ab\to cd}^{uu}(P_T,x_1,x_2) e^{-S_{\rm Sud}^T(Q^2,b_\perp)} \ ,
\end{eqnarray}
where $\mu_b=b_0/b_\perp$ and $S_{\rm Sud}^T(Q^2,b_\perp)$ was defined in Eq.~(\ref{sudakov-spin}). For the unpolarized case, we could in principle also use a more advanced resummed result as derived in Ref.~\cite{Sun:2015doa}. For the following studies, we have decided to use the same resummation accuracy for both the polarized and unpolarized cross sections. We have checked numerically
that this does not introduce any sizable effects for the unpolarized case as far as the spin asymmetries are concerned. 

In order to compare to the experimental data, we have to make further approximations. This is because the phenomenology of the Sivers function (or the Qiu-Sterman matrix element) in the spin-dependent cross section is presently not sophisticated enough to warrant the inclusion of detailed evolution effects in our calculations. Therefore, we will apply estimates of the quark Sivers functions from known experiments without considering their scale dependence. We approximate the parton distributions in the above equations at a fixed lower scale, e.g., $\mu_b\approx \mu_0=2 \rm GeV$, where the quark Sivers functions are extracted from SIDIS. We comment on the uncertainties of these extractions below.  

With these modifications, the resummation formulas can be written as
\begin{eqnarray}
W_{ab\to cd}^{T\beta}|_{\rm MLLA'}&=&\frac{ib_\perp^\beta}{2}x_1 f_a(x_1,\mu_0)x_2 T_{Fb}(x_2,\mu_0) H_{ab\to cd}^{\rm Sivers}(P_T,x_1,x_2) e^{-S_{\rm Sud}^T(Q^2,b_\perp)} \ ,\\
W_{ab\to cd}^{uu}|_{\rm MLLA'}&=&x_1 f_a(x_1,\mu_0)x_2 f_{b}(x_2,\mu_0) H_{ab\to cd}^{uu}(P_T,x_1,x_2) e^{-S_{\rm Sud}^T(Q^2,b_\perp)} \ ,
\end{eqnarray}
where $\mu_0$ will be varied around $2~\rm GeV$ in our final results. In addition, these modifications also allow us to simplify the numerical calculations. Within the above approximation, the $q_\perp$-dependence only comes from the Sudakov form factor (with the additional $b_\perp^\beta$ for the Sivers effect) which also depends on $Q^2$.

\subsection{Preliminary results}

According to the above results, the Sivers asymmetries depend on three ingredients: (1) the Sivers functions; (2)  the associated hard factors for the different partonic channels; (3) the $q_\perp$ dependence from the Sudakov form factor and the related non-perturbative TMDs. In the following, we will briefly go through these three ingredients, and we then address the comparison to the STAR data~\cite{Abelev:2007ii,starpreliminary}.

\subsubsection{Sivers Functions}

The quark Sivers functions have been mainly determined from single transverse spin asymmetries in semi-inclusive hadron production in the deep-inelastic scattering process. The latest global analyses can be found in Refs.~\cite{Cammarota:2020qcw,Bacchetta:2020gko} (see also references therein). The associated Qiu-Sterman matrix elements $T_{F}(x,x)$ have also been extracted from the single spin asymmetry $A_N$ for single inclusive hadron production in $pp$ collisions~\cite{Kouvaris:2006zy}. However, the latter process contains the Collins twist-three fragmentation contributions as well~\cite{Kang:2010zzb,Metz:2012ct,Kanazawa:2014dca,Kanazawa:2015ajw,Gamberg:2017gle}. Therefore, it may not be sufficient to constrain the Sivers functions from inclusive hadron $A_N$ in $pp$ collisions alone. Recently a first attempt was made to perform a global analysis of SSA data from different processes including $A_N$ in $pp$ collisions~\cite{Cammarota:2020qcw}. 

\begin{figure}[tbp]
\centering
\includegraphics[width=9cm]{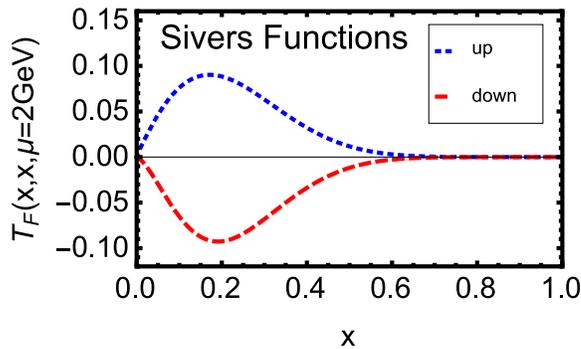}
\caption{{\it The Sivers functions used in our calculations. We use the extractions from Ref.~\cite{Sun:2013hua} which are obtained from Sivers SSAs in semi-inclusive hadron production in deep-inelastic scattering.}}
\label{siversfunction}
\end{figure}

In the following calculations, we will use the Sivers functions constrained by SSA data in SIDIS. We would like to add a few comments concerning the precision of these extractions. First, the current experimental data on SSAs in SIDIS come from the relatively low $Q^2$ region, where TMD factorization may not be as rigorous as compared to high $Q^2$. This situation will of course be improved by the future Electron-Ion Collider. Until then, we have to keep in mind the systematic uncertainties of the Sivers functions extracted from the existing SIDIS data. A crosscheck with other processes, such as Drell-Yan lepton pair production and $W$/$Z$-boson production in $pp$ collisions will provide crucial information on the Sivers functions. Second, the quark Sivers functions determined from SIDIS have a particular feature -- the up quark and down quark distributions have opposite signs. As a result, one often finds a strong cancelation between these two quark Sivers functions
when they are used in a physical process, resulting in significant uncertainties of phenomenological extractions of the functions. All existing fits report at least $10\%$ uncertainty of the valence quark Sivers functions. In addition, the sea quark Sivers functions have even larger uncertainties. We hope that the future EIC will help to better constrain both valence and sea quark Sivers functions. 

For our numerical results we will use the quark Sivers function from Ref.~\cite{Sun:2013hua} as a baseline, keeping in mind their significant uncertainty just described. The dijet asymmetries studied in this paper are precisely a case where the $u$ and $d$ quark Sivers functions are added (at least for the dominant $qg$ channel) and cancel to a significant extent. This also increases the uncertainty of any predictions that can be made. Given the uncertainty of the $u+d$ combination, and to obtain a conservative order of magnitude estimate, we assume the total $u+d$ distribution to be $\pm 0.2u$, allowing the distribution to have either sign. We also neglect the sea quark Sivers function contributions in the numeric estimate. Because of the $qg\to qg$ channel dominance, the individual difference between the $u$ and $d$ quark Sivers in our parameterization will not affect much the final results.

More recently, the STAR collaboration has developed a novel approach to study the SSA in dijet production at RHIC using quark flavor tagging by measuring the charge of jets~\cite{starpreliminary}. By tagging the total charge of the final state jet, one can separate $u$-quark jets from $d$-quark jets which potentially avoids the cancelation between the two distributions to some degree. The SSAs for the flavor tagged dijet production will be directly related to either the $u$-quark Sivers function or the $d$-quark Sivers function, depending on the charge of the triggered jet. We will comment on the comparison to these exciting new data at the end of this section. See also Ref.~\cite{Field:1977fa} for the original proposal of the jet charge by Feynman and Field, Refs.~\cite{Krohn:2012fg,Waalewijn:2012sv} for recent theoretical studies and Refs.~\cite{Aad:2015cua,Sirunyan:2017tyr} for experimental measurements by ATLAS and CMS.

\subsubsection{Hard Factors}

We expect that the quark-gluon channel will make the dominant contribution to the SSA in dijet production, especially
at forward rapidity where one probes the valence region of the Sivers functions, which in fact is primarily constrained
by the SIDIS experiments. It is instructive to examine the hard factor for this channel as an example to study the kinematic dependence of the SSA of our process.

\begin{figure}[tbp]
\centering
\includegraphics[width=9cm]{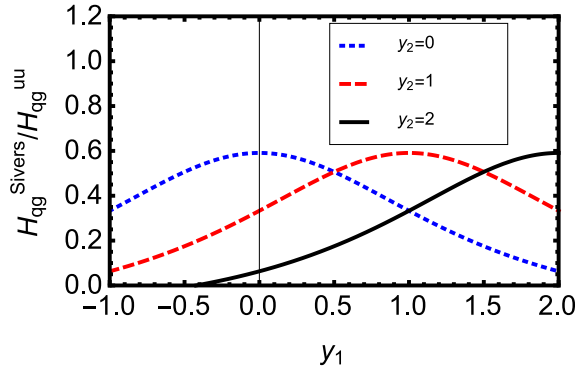}
\caption{{\it Ratio of the hard factors of the Sivers SSA and the spin-averaged cross section as a function of the leading jet's rapidity $y_1$ for typical kinematics of $P_T=6~\rm GeV$ and $y_2=0,\, 1,\,2$, respectively.}}
\label{hardfactor}
\end{figure}

In Fig.~\ref{hardfactor}, we show the ratio of the spin-dependent hard factor and the spin-averaged hard factor for the $qg\to qg$ channel for typical kinematics at RHIC. The leading jet transverse momentum is chosen as $P_T=6\rm$~GeV and the rapidity interval we consider is $-1<y_1<2$. From the figure, we can clearly see that the ratio increases with rapidity. In particular, the asymmetry will be larger in the forward region. 

Another important feature of the hard factor is that it is positive for all kinematics. This means that the asymmetry is dominated by final state interaction effects. This is different from the inclusive hadron $A_N$, which is dominated by initial state interaction effects. The reason is that for dijet production, the final state interaction with the gluon line in the $qg\to qg$ channel cancels the initial state interaction with the initial gluon line. On the other hand, for single inclusive hadron production, the final state interaction with the gluon line does not contribute to the quark fragmentation part in the $qg\to qg$ channel which is the dominant source for hadron production in $pp$ collisions. Therefore, the dijet SSA will have an opposite sign compared to the single inclusive hadron SSA. This is a very interesting feature and will have significant phenomenological consequences for SSAs at RHIC.

\subsubsection{Sudakov Effects}

It has been known for some time that the Sudakov effects will suppress the single spin asymmetries~\cite{Boer:2001he}. This suppression factor was also included when the dijet SSA was proposed in Ref.~\cite{Boer:2003tx}. Here, we would like to follow up on this issue and discuss the effect on the dijet asymmetries using the updated results for the Sudakov form factor and non-perturbative TMDs. We will continue to focus on the $qg$ process.

As mentioned at the beginning of this section, the $q_\perp$-dependence is contained in the Sudakov form factor and the associated non-perturbative TMDs. For the discussion here and the numerical results, we separate the $q_\perp$-dependence of the spin-dependent and spin-averaged differential cross sections as
\begin{eqnarray}
{\cal R}^{(qg)}(q_\perp)&=&\frac{1}{2\pi}\int db_\perp^2 J_0(q_\perp b_\perp)\, e^{-S_{\textrm{pert}}^{(qg)}(Q^2,b_*)-S_{\textrm{NP}}^{(qg)}(Q,b_\perp)} \ ,\\
{\cal R}_T^{(qg)}(q_\perp)&=&\frac{1}{4\pi}\int db_\perp^2 J_1(q_\perp b_\perp)\, e^{-S_{\textrm{pert}}^{(qg)}(Q^2,b_*)-S_{\textrm{NP}}^{T(qg)}(Q,b_\perp)} \ ,
\end{eqnarray}
where $J_{0,1}$ are Bessel functions and where we have applied the $b_*$-prescription in the above equation: $b_*=b_\perp/\sqrt{1-b_\perp^2/b_{{\textrm{max}}}^2}$ with $b_{{\textrm{max}}}=2~\rm GeV^{-2}$. The perturbative part of the Sudakov form factor is the same for both cases $S_{\textrm{pert}}^{(qg)}(Q^2,b_*)=S_{\textrm{Sud}}^{T(qg)}(Q^2,b_*)$ which was given in Eq.~(\ref{sudakov-spin}). The non-perturbative parts are parametrized as~\cite{Su:2014wpa}
\begin{eqnarray}
S_{\textrm{NP}}^{(qg)}(Q,b_\perp)&=&\left(C_F+C_A\right)\left[\frac{g_1}{2} b_\perp^2 + \frac{g_2}{2} \ln \frac{Q}{Q_0} \ln \frac{b_\perp}{b_*}\right]\ ,\\
S_{\textrm{NP}}^{T(qg)}(Q,b_\perp)&=&S_{\textrm{NP}}^{(qg)}(Q,b_\perp)-g_s b^2 \ ,
\end{eqnarray}
with $g_1 = 0.212$, $g_2 = 0.84$, $g_s=0.062$ and $Q_0^2 = 2.4$ GeV$^2$. 

\begin{figure}[tbp]
\centering
\includegraphics[width=9cm]{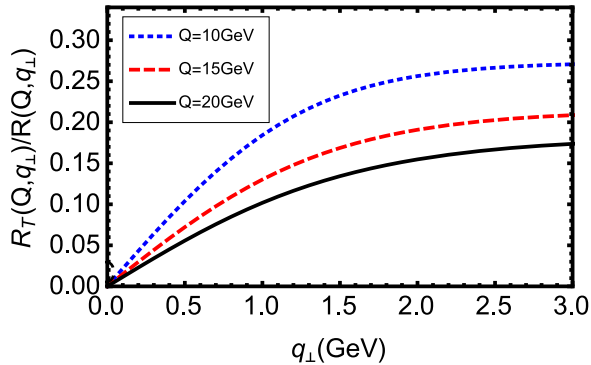}
\caption{{\it The ratio of ${\cal R}_T(Q,q_\perp)$ and ${\cal R}(Q,q_\perp)$ as a function of $q_\perp$ for typical $Q$ values to dijet production at RHIC.}}
\label{sudakovqt}
\end{figure}

As an example, we plot in Fig.~\ref{sudakovqt} the ratio ${\cal R}_T/{\cal R}$ as a function of $q_\perp$ for typical values of $Q$ relevant for RHIC kinematics. Clearly, the asymmetry increases with $q_\perp$. Different from previous examples (SIDIS or Drell-Yan lepton pair production), the curves in the plot do not reach the maximum of the asymmetry in the $q_\perp$ region relevant for TMD physics. The reason is that for the $qg$ channel, the Sudakov form factor leads to a strong $q_\perp$-broadening effect. In particular, this is due to the the double logarithms associated with the incoming gluon distribution which push the peak of the asymmetry to higher values of $q_\perp$, beyond the TMD domain. 

\subsection{Comparison to the STAR Data from 2007}

As suggested in Ref.~\cite{Boer:2003tx}, the dijet spin asymmetry can be measured through the azimuthal angular distribution between the two jets. Because the Sivers asymmetry leads to a preferred direction of the total transverse momentum of the two final state jets, the angular distribution will be shifted toward that direction related to the Sivers asymmetry. The magnitude of the asymmetry will depend on the relative angle between the leading jet and the polarization vector $\vec{S}_\perp$. In particular, as mentioned in Ref.~\cite{Boer:2003tx}, the SSA for dijet production is at its maximum value when the jet direction is parallel or anti-parallel to the spin vector $\vec{S}_\perp$ of the proton. However, the asymmetry will vanish if the leading jet is perpendicular to the spin $\vec{S}_\perp$. It can be shown that this introduces an additional factor of $|\cos(\phi_j-\phi_S)|$ for the dijet SSA. Therefore, when we compare to the STAR data, we need to include an average of this factor over the azimuthal angle between the leading jet and the polarization vector: $\frac{1}{\pi}\int_0^\pi d\phi |\cos(\phi)|={2\over \pi}$.

As mentioned above, for the dijet SSA we take the $u+d$ Sivers distribution to be $20\%$ of the extracted $u$-quark Sivers function. From the existing experimental data, we can determine neither the size nor the sign of the total contribution from the $u$ and $d$-quark Sivers functions. Therefore, 
to estimate the total contributions to the dijet SSA, we will use
both signs, in order to obtain an idea of the uncertainties associated with these determinations. 

The jet kinematics of the data published by the STAR experiment in 2007 is $P_T>4~\rm GeV$ and rapidity in the range of $-1<y_{1,2}<2$. In order to compare to the experimental data, we integrate out the leading jet momentum and the relative rapidity between the two jets, but we keep the total rapidity $y=y_1+y_2$. 

\begin{figure}[tbp]
\centering
\includegraphics[width=9cm]{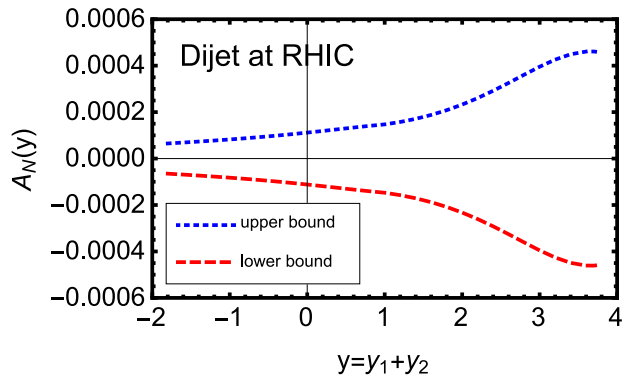}
\caption{{\it The SSA in dijet production at RHIC as a function of the total rapidity $y=y_1+y_2$ of the two jets for the kinematics of the STAR measurement in 2007: $P_T>4~$GeV and $-1<y_{1,2}<2$. The upper bound corresponds to $T_{Fu}+T_{Fd}$ with $+20\%$ of the fitted value of $T_{Fu}$ of Ref.~\cite{Sun:2013hua}, whereas the lower bound corresponds to $-20\%$.}}
\label{star2007}
\end{figure}

Using the differential cross sections for the spin-averaged and spin-dependent cases in the MLLA given above, we obtain the following expression for the single transverse spin asymmetry which can be compared to the STAR measurement:
\begin{eqnarray}
A_N(y)=\frac{2}{\pi}\frac{\sum_{acd}\sum_{b=u,d}\int d^2P_Tdy_1 \,x_1f_a(x_1,\mu_0)\,x_2T_{Fb}(x_2,\mu_0)\,H_{ab\to cd}^{\rm Sivers}(P_T,x_1,x_2)\, {w}_T(Q)}{\sum_{abcd} \int d^2P_Tdy_1 \,x_1f_a(x_1,\mu_0)\,x_2f_{b}(x_2,\mu_0)\,H_{ab\to cd}^{uu}(P_T,x_1,x_2) \,{w}(Q)} \ ,\ \ \ \ \label{anystar2007}
\end{eqnarray}
where $Q^2=\hat s=x_1x_2S_{pp}$ and $w(Q),w_T(Q)$ are weights for the total transverse momentum integral,
\begin{eqnarray}
w_T(Q)&=&\int_0^{Q/6} dq_\perp q_\perp {\cal R}_T(Q,q_\perp) \ , \\
w(Q)&=&\int_0^{Q/6} dq_\perp q_\perp {\cal R}(Q,q_\perp) \ .
\end{eqnarray}
The upper limits of the above integrals correspond to the TMD region where $q_\perp\ll P_T$. Notice that $Q\ge 2P_T$ for all kinematics. 

With our assumptions on the $u$ and $d$ Sivers functions, we calculate the SSA in Eq.~(\ref{anystar2007}), and find that the asymmetry is of order $10^{-4}$ for the entire rapidity range shown, see, Fig.~\ref{star2007}. We note that the central values of $u$-quark and $d$-quark Sivers functions from the fit of Ref.~\cite{Sun:2013hua} lead to smaller asymmetries shown in Fig.~\ref{star2007}. Let us summarize the main differences with respect to the previous calculation in Ref.~\cite{Bomhof:2007su}: First, the Sivers functions determined from SIDIS are different. Second, we have included the relative suppression from Sudakov effects. If we integrate over the entire rapidity region, the asymmetry is about $1.7\times 10^{-4}$.

It is interesting to note that the STAR measurement in 2007 found that the SSA for dijet production is consistent with zero, and their uncertainties are of the order of $10^{-2}$. According to our calculation, a finite asymmetry could be observed if the uncertainty is reduced by more than one order of magnitude. Of course, this also depends on the size of the total up and down quark Sivers function. If they completely cancel, then the asymmetries will depend on the sea quark Sivers functions, which are known to be not as large as the valence ones.

\subsection{The Flavor Tagged Dijet Asymmetry}

In the last few years, the STAR collaboration has investigated a novel method to explore the SSA in dijet production. They considered the jet charge to tag the up or down quark flavor of the jet. An up (down) quark jet has positive (negative) jet charge, whereas a gluon jet leads to a neutral jet charge. From the preliminary analysis, the efficiency of this separation is reasonably good. A similar idea is also proposed in~\cite{Kang:2020fka}. This also suggests further improvements by using the jet flavor information in the jet charge definition.

\begin{figure}[tbp]
\centering
\includegraphics[width=9cm]{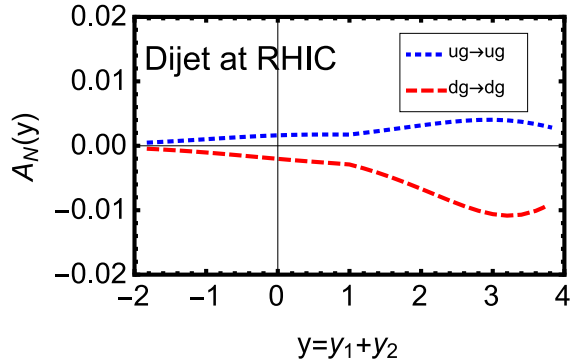}
\caption{{\it The SSA in dijet production for the $qg\to qg$ channel only. We show the result separately for up and down quarks.}}
\label{flavortag}
\end{figure}

In order to compare to the experimental data, we take into account the $u$ and $d$ quark contributions separately, just for the $qg\to qg$ channel. For up quarks, we have
\begin{eqnarray}
A_N^{{\textrm{(up)}}}(y)=\frac{2}{\pi}\frac{\int d^2P_Tdy_1\, x_1f_g(x_1,\mu_0)\,x_2T_{Fu}(x_2,\mu_0)\,H_{gq\to gq}^{\rm Sivers}(P_T,x_1,x_2)\, {w}_T(Q)}{ \int d^2P_Tdy_1 \left[ x_1f_g(x_1,\mu_0)\,x_2f_{u}(x_2,\mu_0)\,H_{gq\to gq}^{uu}(P_T,x_1,x_2)+(x_1\leftrightarrow x_2)\right] {w}(Q)} \ . \ \ \  \ \label{asymmetryup}
\end{eqnarray}
An analogous expression holds for the $d$-quark Sivers contribution. In Fig.~\ref{flavortag}, we plot the two asymmetries as functions of $y=y_1+y_2$. From this plot, we can see that the asymmetries are of the order of $10^{-3}$. If we integrate over the entire rapidity range, we obtain an asymmetry of $2.2\times 10^{-3}$ and $-3.7\times 10^{-3}$ for the $ug\to ug$ and $dg\to dg$ channels, respectively. Our results compare reasonably well to the preliminary STAR results~\footnote{An estimate from the preliminary STAR analysis in Ref.~\cite{starpreliminary} suggests that the observed asymmetry is around $1.5\times 10^{-3}$ and $-1.8\times 10^{-3}$ for positively and and negatively tagged charged jets, respectively. We hope that the spin asymmetry defined in Eq.~(\ref{asymmetryup}) will be reported in the new analysis as well.}. 

Compared to the results shown in Fig.~\ref{star2007}, we find that the asymmetries are much larger for the flavor tagged case. This is not only due to the fact that for the flavor tagged case no cancelation between the $u$ and $d$-quark Sivers functions occurs, but also because the denominator of the unpolarized cross section only contains the $qg\to qg$ channel. By tagging the (quark) flavor of the final state jets, we exclude a major background contribution from $gg\to gg$, which only leads to charge neutral jets in the final state.

The asymmetries shown in Fig.~\ref{flavortag} assume $100\%$ efficiency of the tagging in the jet sample. To compare to the STAR data properly, we need to consider a realistic tagging efficiency, which will suppress the asymmetries somewhat. 

\section{Summary and Discussion}

We have investigated the single transverse spin asymmetry in dijet correlations in hadronic collisions. The total transverse momentum of the dijet in the final state is correlated with the incoming nucleon's polarization vector. We have focused on the SSA contribution from the quark Sivers function of the polarized nucleon where initial and final state interaction effects play an important role. A detailed analysis at one-loop order has been carried out for the contribution from soft gluon emissions in order to understand the factorization properties. It was found that the associated TMD factorization is valid at the level of leading double logarithms and single logarithms from the TMD quark distribution and those collinear to the jet. However, additional soft gluon contributions to the single logarithms do not show the same pattern as in the unpolarized case investigated in Ref.~\cite{Sun:2015doa} and hence cannot be incorporated in the TMD factorization formula in terms of the same spin independent soft factor. This indicates that the standard TMD factorization is broken at the single logarithmic level for the SSA in dijet correlations in hadronic collisions. We believe that this issue will deserve further attention by investigating whether a consistent ``non-traditional'' soft factor for the single transverse spin asymmetry could be defined. 

We have further presented phenomenological studies for the SSA in dijet production at RHIC based on the LLA$'$ approximation, for which one improves the standard LLA resummation formula by ``universal'' subleading logarithmic terms associated with the initial partons and the final state jets. Using the quark Sivers functions constrained by SSAs in SIDIS, we have found that the leading double logarithmic resummation effects suppress the asymmetry for the kinematics relevant for the measurements by the STAR collaboration at RHIC, making them broadly consistent with experimental results. We also presented results for the flavor (charge) tagged dijet case, where the asymmetries are much larger than when the jet flavor is not tagged. A detailed comparison with the experimental data will be helpful to understand factorization breaking effects.

We finally note that in our analysis we have only considered the perturbative part of the factorization breaking effects. The non-perturbative TMD quark distribution could also contribute to such effects. Our numerical results assume that this part is the same as for SIDIS. To address this issue, more detailed comparisons to experimental measurements and further theoretical efforts are necessary. In any case, a more refined phenomenology will be warranted once there is a better general understanding of factorization breaking effects, along with more detailed experimental information.

\vspace{2em}

\section*{Acknowledgement}

This work is partially supported by the U.S. Department of Energy, Office of Science, Office of Nuclear Physics under contract number DE-AC02-05CH11231. This study was supported by Deutsche Forschungsgemeinschaft (DFG) through the Research Unit FOR 2926 (project number 40824754).

\appendix

\section{Soft Gluon Radiation Associated with the Jet}

In this section, we derive the soft gluon radiation contribution associated with the final state jets. Following Ref.~\cite{Liu:2018trl,Liu:2020dct}, we apply the following identity:
\begin{eqnarray}
S_g(k_1,p_1)+S_g(k_1,p_2)&=&\frac{4}{k_{g\perp}^2}+\frac{4}{k_{g\perp}^2}\frac{\vec{k}_{1\perp}\cdot \vec{k}_{g\perp}}{k_1\cdot k_g} \ ,
\end{eqnarray}
where the first term contributes to the double logarithms and corresponds to the first term in the bracket of Eq.~(\ref{k1p1}). To calculate the second term, we define 
\begin{equation}
I(R_1)=\int\frac{d\xi}{\xi}\frac{d\phi}{2\pi}\frac{\vec{k}_{1\perp}\cdot \vec{k}_{g\perp}}{k_1\cdot k_g} \ ,
\end{equation}
where $\xi$ is the longitudinal momentum of $k_g$ with respect to $k_1$, i.e. $\xi=k_g\cdot p_1/k_1\cdot p_1$ and $\phi$ is the azimuthal angle between $k_{g\perp}$ and $k_{1\perp}$. This integral was calculated in Refs.~\cite{Liu:2018trl,Liu:2020dct}, and we find
\begin{equation}
I(R)=-\frac{1}{\epsilon}\left[1-R^{-2\epsilon}\right]=\ln\frac{1}{R^2}+\epsilon\frac{1}{2}\ln^2\frac{1}{R^2} \ .
\end{equation}
Similarly, we find
\begin{eqnarray}
S_g(k_1,p_1)-S_g(k_1,p_2)
&=&\frac{4}{k_{g\perp}^2}\frac{k_1^+k_g^--k_1^-k_g^+}{k_1\cdot k_g}  \ .
\end{eqnarray}
We notice that there is no divergence associated with the jet in the above term. After averaging over the azimuthal angle of the jet, we have
\begin{eqnarray}
\int\frac{d\xi}{\xi}\left[S_g(k_1,p_1)-S_g(k_1,p_2)\right]&=&\frac{4}{k_{g\perp}^2}\int_{\xi_0}^{\xi_1}\frac{d\xi}{\xi}\frac{k_{g\perp}^2-\xi^2k_{1\perp}^2}{|k_{g\perp}^2-\xi^2k_{1\perp}^2|}\nonumber\\
&&\times \left[1-2\epsilon\ln\frac{2|k_{g\perp}^2-\xi^2k_{1\perp}^2|}{k_{g\perp}^2+\xi^2k_{1\perp}^2+|k_{g\perp}^2-\xi^2k_{1\perp}^2|}\right]\ ,
\end{eqnarray}
where the boundary of the $\xi$-integral is determined by the kinematic constraints: $\xi_0=k_{g\perp}^2/(-\hat t)$ and $\xi_1=(-\hat u)/k_{1\perp}^2$. Since there is no divergence resulting from the $\xi$-integral, we obtain the following final result
\begin{eqnarray}
\int\frac{d\xi}{\xi}\left[S_g(k_1,p_1)-S_g(k_1,p_2)\right]&=&\frac{1}{k_{g\perp}^2}\ln\frac{\hat t}{\hat u} \ .
\end{eqnarray}
In particular, we note that the $\epsilon$-term vanishes after the integration. Combining the above equations, we obtain the result in Eqs.~(\ref{e11}), (\ref{e12}). Similarly, we can derive the results in Eqs.~(\ref{e21}), (\ref{e22}). The result for $S_g(k_1,k_2)$ is a little more involved. We notice that
\begin{eqnarray}
S_g(k_1,k_2)=\frac{2k_1\cdot k_2}{k_1\cdot k_g k_2\cdot k_g}=S_g(P,k_1)+S_g(P,k_2) \ ,
\end{eqnarray}
where $P=p_1+p_2=k_1+k_2$. We further subtract the divergence associated with the jet contribution,
\begin{eqnarray}
S_g(P,k_1)-S_g(p_1,k_1)=4\frac{\xi^2k_{1\perp}^2-k_{g\perp}^2}{k_{g\perp}^2+\tilde{x}^2\hat s}\frac{1}{(k_{g\perp}-\xi k_{1\perp})^2}\ ,
\end{eqnarray}
where $\tilde x=k_g\cdot p_1/p_1\cdot p_2$. Clearly the collinear divergence associated with the jet is canceled between these two terms. Averaging over the azimuthal angle of the jet, we have
\begin{eqnarray}
4\int_{\xi_0}^{\xi_1}\frac{d\xi}{\xi}\frac{k_{g\perp}^2-\xi^2k_{1\perp}^2}{|k_{g\perp}^2-\xi^2k_{1\perp}^2|}\frac{1}{k_{g\perp}^2+\tilde x^2\hat s} \left[1-2\epsilon\ln\frac{2|k_{g\perp}^2-\xi^2k_{1\perp}^2|}{k_{g\perp}^2+\xi^2k_{1\perp}^2+|k_{g\perp}^2-\xi^2k_{1\perp}^2|}\right] \ ,
\end{eqnarray}
where $\xi_{0,1}$ are the same as above. Carrying out the integral, we find 
\begin{eqnarray}
-\frac{1}{2k_{g\perp}^2}\left[\ln\frac{\hat s}{k_{g\perp}^2}+\ln\frac{\hat t^2}{\hat s^2}-2\epsilon \ln\frac{\hat s}{-\hat t}\ln\frac{\hat s}{-\hat u}\right] \ .
\end{eqnarray}
Together with the result for $S_g(p_1,k_1)$ we obtain the result for $S_g(P,k_1)$. A similar result can be obtained for $S_g(P,k_2)$, and we are able to derive the result for $S_g(k_1,k_2)$ in Eq.~(\ref{e22p}).

\end{document}